\documentclass[preprint]{elsarticle}
\usepackage[a4paper, total={7in, 8in}]{geometry}
\usepackage[utf8]{inputenc}
\usepackage{floatpag}
\usepackage{amsthm}
\usepackage{enumerate}
\usepackage{amssymb}
\usepackage{amsmath}
\usepackage{braket}
\usepackage{hyperref}
\hypersetup{colorlinks = true, linkcolor=blue }

\usepackage{graphicx}
\usepackage{epstopdf}
\usepackage{color} 
\graphicspath{{./Figures/}}
\usepackage{caption}
\usepackage{subcaption}
\usepackage{cleveref}

\usepackage{etoolbox}
\patchcmd{\pprintMaketitle}
  {\fi\hrule}
  {\fi\ifvoid\extrainfobox\else\unvbox\extrainfobox\par\vskip10pt\fi\hrule}
  {}{}

\newenvironment{extrainfo}
  {\global\setbox\extrainfobox=\vbox\bgroup\parindent=0pt }
  {\egroup}
\newsavebox\extrainfobox

\journal{Physics Letters B}

\begin{document}

\begin{frontmatter}

\begin{extrainfo}
\rightline{\large CERN-TH-2019-102}
\end{extrainfo}

\title{{\bf Effect of non-eikonal corrections on azimuthal asymmetries in the Color Glass Condensate}}

\author[a]{Pedro Agostini}
\ead{pedro.agostini@usc.es}
\author[b]{Tolga Altinoluk}
\ead{Tolga.Altinoluk@ncbj.gov.pl}
\author[a,c]{N\'estor Armesto}
\ead{nestor.armesto@usc.es}

\address[a]{Instituto Galego de F\'{\i}sica de Altas Enerx\'{\i}as IGFAE, Universidade de Santiago de Compostela, 15782 Santiago de Compostela, Galicia-Spain}
\address[b]{National Centre for Nuclear Research, 00-681 Warsaw, Poland}
\address[c]{Theoretical Physics Department, CERN, 1211 Geneva 23, Switzerland}


\begin{abstract}
We analyse the azimuthal structure of two gluon correlations in the Color Glass Condensate including those effects that result from relaxing the shockwave approximation for the target. Working in the Glasma graph approach suitable for collisions between dilute systems, we compute numerically the azimuthal distributions and show that both even and odd harmonics appear. We study their dependence on model parameters, energy of the collision, pseudorapidity and transverse momentum of the produced particles, and length of the target. While the contribution from non-eikonal corrections vanishes with increasing collision energy and becomes negligible at the energies of the Large Hadron Collider, it is found to be sizeable up to top energies at the Relativistic Heavy Ion Collider.
\end{abstract}

\end{frontmatter}

\section{Introduction}
\label{sec:intro}

The existence of azimuthal asymmetries in particle production stretched for a long pseudorapidity interval - named the ridge - has been observed in small collision systems, proton-proton and proton-nucleus, at the Large Hadron Collider (LHC) at CERN~\cite{Khachatryan:2010gv,Khachatryan:2015lva,Aad:2015gqa,CMS:2012qk,Abelev:2012ola,Aad:2012gla,Aaij:2015qcq,Khachatryan:2016ibd,Khachatryan:2016txc,Aaboud:2016yar,Aaboud:2017acw,Aaboud:2017blb,Chatrchyan:2013nka,Abelev:2014mda} and the Relativistic Heavy Ion Collider (RHIC) at BNL~\cite{Alver:2009id,Abelev:2009af,Adare:2014keg,Adamczyk:2015xjc,Adare:2015ctn}. The corresponding observation in nucleus-nucleus collisions finds a standard explanation in final state interactions that lead to a macroscopic description in terms of relativistic hydrodynamics. But for small systems such explanation is a matter of active debate, see e.g. the recent works \cite{Blok:2018xes,Kurkela:2019kip,Nie:2019swk}, and initial state dynamics have also been invoked.

Concerning initial state explanations, those based on the effective theory for high-energy Quantum Chromodynamics~\cite{Kovchegov:2012mbw} named the Color Glass Condensate (CGC)~\cite{Iancu:2002xk,McLerran:2008uj,Gelis:2010nm}, have been explored intensively in recent years. For the collision of dilute objects like proton-proton, the ``Glasma graph" approximation~\cite{Dumitru:2008wn,Dumitru:2010iy}, that encodes both Bose enhancement and Hanbury-Brown--Twiss (HBT) effects~\cite{Kovchegov:2012nd,Kovchegov:2013ewa,Altinoluk:2015uaa,Altinoluk:2015eka}, has been developed and used to describe experimental data~\cite{Dusling:2012iga,Dusling:2012cg,Dusling:2012wy,Dusling:2013qoz}. Besides two gluon correlations, those among three and four~\cite{Ozonder:2014sra,Ozonder:2017wmh} have also been studied, and also those between two quarks~\cite{Altinoluk:2016vax,Martinez:2018ygo}. The extension to dilute-dense (proton-nucleus) collisions was later done numerically~\cite{Lappi:2015vta} and analytically~\cite{Altinoluk:2018hcu,Altinoluk:2018ogz,Martinez:2018tuf}, including three gluon correlations~\cite{Altinoluk:2018ogz}, and applied to describe data~\cite{Dusling:2017dqg,Dusling:2017aot}. Complementary explanations in terms of density gradients~\cite{Levin:2011fb} have also been considered to explain the observed azimuthal structure. 

In this framework, the two remaining key problems are  the analytical extension to dense-dense collisions, and the absence of odd azimuthal harmonics in standard calculations. To overcome the latter,  several alternatives have been essayed: density corrections in the projectile~\cite{McLerran:2016snu,Kovner:2016jfp,Kovchegov:2018jun} (implemented to attempt a description of data in~\cite{Mace:2018vwq,Mace:2018yvl,Mace:2019rtt}), quark correlations~\cite{Dumitru:2014vka,Kovner:2017gab,Dusling:2017dqg,Davy:2018hsl} and a more involved description of the target~\cite{Kovner:2012jm,Dumitru:2014yza} than the one provided by the commonly used McLerran-Venugopalan (MV) model~\cite{McLerran:1993ni,McLerran:1994vd}.

In this work we explore a different direction. Usual calculations in the CGC employ  the eikonal approximation: the process of propagation of an energetic parton from the projectile through the target, considered as a background field, is computed in the light cone gauge neglecting its transverse components and considering it as infinitely time dilated and Lorentz contracted -- a shockwave. Also terms subleading in energy (among them, spin flip ones) are disregarded. This is to be contrasted to the calculations of elastic and radiative energy loss of energetic partons traversing a medium composed of coloured scattering centers -- jet quenching. Here, the shockwave approximation is relaxed and the target is considered to have a finite length, see e.g. the reviews \cite{Kovner:2003zj,CasalderreySolana:2007pr}\footnote{Non-eikonal corrections at high energies have also been analysed in relation to Transverse Momentum Distributions and spin physics~~\cite{Balitsky:2015qba,Balitsky:2016dgz,Kovchegov:2015pbl,Kovchegov:2016zex,Kovchegov:2017jxc,Kovchegov:2017lsr,Chirilli:2018kkw}, and soft gluon exponentiation~\cite{Laenen:2008ux,Laenen:2008gt,Laenen:2010uz}.}.

Some years ago, a systematic expansion of the gluon propagator in non-eikonal terms stemming from the relaxation of the shockwave approximation was performed in~\cite{Altinoluk:2014oxa,Altinoluk:2015gia} and applied to particle production in the CGC in~\cite{Altinoluk:2015xuy}. Using those ideas, in a recent paper~\cite{Agostini:2019avp} we have computed single, double and triple gluon production in the CGC including those non-eikonal corrections  within the Glasma graph approximation -- thus suitable for collisions of two dilute objects. It was anticipated there that an asymmetry between the near and away side ridges appeared for certain kinematic regions, which would lead to odd azimuthal harmonics. Restricted to two gluon correlations, it is the goal of the present work to study numerically the impact of these non-eikonal corrections on even and odd harmonics, and their dependence on model parameters, energy of the collision, pseudorapidity and transverse momentum of the produced particles, and length of the target.

As discussed in this introduction, non-eikonal corrections are not the only source of odd harmonics, others being density corrections or a more sophisticated treatment of the target beyond the MV model. Besides, they vanish with increasing energy, a trend that is not observed for the odd harmonics in experimental data. Therefore, here we make no attempt to compare with experimental data but only address the existence and size of the non-eikonal effects on the azimuthal structure.

The plan of the paper is as follows: In Section \ref{sec:correl} we present the formulae for two-gluon correlations in a form derived from that in~\cite{Agostini:2019avp} but more suitable for a numerical implementation, and present the details of the model. In Section \ref{sec:azimuth} we show the results for azimuthal harmonics. Finally, in Section \ref{sec:conclu} we provide our conclusions and outlook.

\section{Non-eikonal double gluon production}
\label{sec:correl}

As shown in~\cite{Agostini:2019avp}, the inclusive cross section for the production of two gluons with transverse momenta $\textbf{k}_1$ and $\textbf{k}_2$, and rapidities with $\eta_1$ and $\eta_2$, can be written as
\begin{align}\label{eqn1}
    \frac{d \sigma}{d^2 \textbf{k}_1 d\eta_1 d^2\textbf{k}_2 d\eta_2}&=2 \, (4\pi)^2 \, \alpha_s^2 \, g^4 \, C_A^2 \, (N_c^2-1)^2 \, 
    {\cal G}_1^{\rm NE}(k_1^-;\lambda^+) \,  {\cal G}_1^{\rm NE}(k_1^-;\lambda^+)
    \int_{\textbf{q}_1 \textbf{q}_2}  |a(\textbf{q}_1)|^2 |a(\textbf{q}_2)|^2  \nonumber \\
&\times \bigg\{ I_{\rm 2tr}^{(0)}+\frac{1}{N_c^2-1} \left[  I_{\rm 2tr}^{(1)}+ I_{\rm 1tr}^{(1)} \right] \bigg\},
\end{align}
where
\begin{align}\label{eqn2}
I_{\rm 2tr}^{(0)}= \mu^2\big[\textbf{k}_1-\textbf{q}_1,\textbf{q}_1-\textbf{k}_1\big] \,  
\mu^2\big[\textbf{k}_2-\textbf{q}_2,\textbf{q}_2-\textbf{k}_2\big] 
 L^i(\textbf{k}_1,\textbf{q}_1) L^i(\textbf{k}_1,\textbf{q}_1) \, L^j(\textbf{k}_2,\textbf{q}_2) L^j(\textbf{k}_2,\textbf{q}_2),
\end{align}
\begin{align}\label{eqn3}
I_{\rm 2tr}^{(1)}&= \Big\{{\cal G}_2^{\rm NE}(k_1^-,k_2^-; L^+) \,  
\mu^2\big[\textbf{k}_1-\textbf{q}_1,\textbf{q}_2-\textbf{k}_1\big] \, 
\mu^2\big[\textbf{k}_2-\textbf{q}_2,\textbf{q}_1-\textbf{k}_2\big] \nonumber\\
& \hspace{0.5cm}\times 
\,  L^i(\textbf{k}_1,\textbf{q}_1) L^i(\textbf{k}_1,\textbf{q}_2) \, L^j(\textbf{k}_2,\textbf{q}_2) L^j(\textbf{k}_2,\textbf{q}_1) \Big\}
+(\underline{k}_2\rightarrow -\underline{k}_2)
\end{align}
and
\begin{align}\label{eqn4}
I_{\rm 1tr}^{(1)}&= \bigg\{ \mu^2\big[\textbf{k}_1-\textbf{q}_1,\textbf{q}_2-\textbf{k}_2\big] \, 
 \mu^2\big[ \textbf{k}_2-\textbf{q}_2,\textbf{q}_1-\textbf{k}_1\big] \, 
 L^i(\textbf{k}_1,\textbf{q}_1) L^i(\textbf{k}_1,\textbf{q}_1) \,
  L^j(\textbf{k}_2,\textbf{q}_2) L^j(\textbf{k}_2,\textbf{q}_2) \nonumber \\
&+{\cal G}_2^{\rm NE}(k_1^-,k_2^-; L^+)\bigg\lgroup
\mu^2\big[\textbf{k}_1-\textbf{q}_1,\textbf{q}_1-\textbf{k}_2\big] \,
\mu^2\big[\textbf{k}_2-\textbf{q}_2,\textbf{q}_2-\textbf{k}_1)\big] 
  + 
  \frac{1}{2}
   \mu^2\big[ \textbf{k}_1-\textbf{q}_1,\textbf{k}_2-\textbf{q}_2\big] \, 
   \mu^2\big[\textbf{q}_2-\textbf{k}_1,\textbf{q}_1-\textbf{k}_2\big] \bigg\rgroup\nonumber \\
&\times L^i(\textbf{k}_1,\textbf{q}_1) L^i(\textbf{k}_1,\textbf{q}_2) \, L^j(\textbf{k}_2,\textbf{q}_1) L^j(\textbf{k}_2,\textbf{q}_2) \bigg\}
+(\underline{k}_2\rightarrow -\underline{k}_2).
\end{align}
For the sake consistency, we use the same notation that was introduced in \cite{Agostini:2019avp}. The subscripts on the right hand side of Eqs.{\eqref{eqn2}}, {\eqref{eqn3}}   and {\eqref{eqn4}} stands for single and double trace operators. These originate from the weak field expansion of the double dipole and quadrupole operators that are present in the production cross section in pA collisions (see \cite{Agostini:2019avp} for details).    Here, we work in light-cone coordinates $(a^+,a^-,\textbf{a})$, superindices $i$ denote transverse coordinates, 
we use the shorthand notation $\underline{k} \equiv (k^+,\textbf{k})$ for the three-momenta of the produced gluons, $\int_{\textbf{q}}\equiv \int d^2\textbf{q}/(2\pi)^2$,  $N_c$ is the number of colors, $\alpha_s=g^2/(4 \pi)$ the strong coupling constant, and the non-eikonal correction functions coming from the finite extension of the target in the $+$ lightcone direction $L^+$ read
\begin{align}
\label{eq:c1}
{\cal G}_1^{\rm NE}(k^-;\lambda^+)= \frac{1}{k^- \lambda^+} \sin (k^- \lambda^+)
\end{align}
and
\begin{align}\label{eqn6}
{\cal G} _2^{\rm NE}(k_1^-,k_2^-; L^+)=\Bigg\{ \frac{2}{\left(k_1^--k_2^-\right)L^+} \sin \bigg[  \frac{(k_1^--k_2^-)}{2} L^+ \bigg]\Bigg\}^2
\end{align}
with $\lambda^+ \ll L^+$ the correlation length of the color sources in the target and $k^-=\textbf{k}^2/2k^+$. Function $\mu^2(\textbf{k},\textbf{q})$ denotes the Fourier transform of the averages of the color charge distributions in the projectile,
\begin{equation}
\label{lipatov}
L^i(\textbf{k}_1,\textbf{q}_1)=\bigg[\frac{(\textbf{k}_1-\textbf{q}_1)^i}{(\textbf{k}_1-\textbf{q}_1)^2}-\frac{\textbf{k}_1^i}{\textbf{k}_1^2}\bigg]
\end{equation} is the usual eikonal Lipatov vertex and function $a(\textbf{q})$ is the functional form of the target potential in momentum space that appears in the definition of the average of the two target field correlator (see \cite{Agostini:2019avp} for the details of the set up and the derivation of the double inclusive gluon production cross section).

To evaluate \cref{eqn1}, we make some assumptions:
\begin{enumerate}
	\item We assume a Gaussian distribution of the colour sources inside the projectile, the MV model~\cite{McLerran:1993ni,McLerran:1994vd}, such that
	\begin{align}
	\mu^2(\textbf{k},\textbf{q})=\mu^2 (2 \pi)^2 \delta^{(2)}(\textbf{k}+\textbf{q}),
	\end{align}
	where $\mu$ is the width of the Gaussian and has units of mass squared.
	
	\item We choose a Yukawa-type potential generated by the colour sources inside the target:
	\begin{align}
	|a(\textbf{q})|^2=\frac{\mu_T^2}{(q^2+\mu_T^2)^2},
	\label{eq:yuk}
	\end{align}
	where $\mu_T$ is an infrared regulator analogous to a Debye mass.
\end{enumerate}

With these assumptions,  \cref{eqn2,eqn1,eqn3} can be further simplified and  their final forms read
\begin{align}\label{eqn9}
I_{\rm 2tr}^{(0)}=\mu^4 S_\perp^2 \,  L^i(\textbf{k}_1,\textbf{q}_1) L^i(\textbf{k}_1,\textbf{q}_1)\,  L^j(\textbf{k}_2,\textbf{q}_2) L^j(\textbf{k}_2,\textbf{q}_2),
\end{align}
\begin{align}\label{eqn10}
I_{\rm 2tr}^{(1)}=\mu^4 S_\perp  \, {\cal G}_2^{\rm NE}(k_1^-,k_2^-; L^+) \, 
\big[ (2\pi)^2 \delta^{(2)}(\textbf{q}_1-\textbf{q}_2) \big]
 L^i(\textbf{k}_1,\textbf{q}_1) L^i(\textbf{k}_1,\textbf{q}_2) \, 
 L^j(\textbf{k}_2,\textbf{q}_2) L^j(\textbf{k}_2,\textbf{q}_1)+(\underline{k}_2\rightarrow -\underline{k}_2)
\end{align}
and
\begin{align}\label{eqn11}
I_{\rm 1tr}^{(1)}&=\mu^4 S_\perp
 \bigg\{
 \Big\lgroup (2\pi)^2 \delta^{(2)}\big[\textbf{k}_1-\textbf{q}_1-(\textbf{k}_2-\textbf{q}_2)\big]\Big\rgroup
  L^i(\textbf{k}_1,\textbf{q}_1) L^i(\textbf{k}_1,\textbf{q}_1) \, 
  L^j(\textbf{k}_2,\textbf{q}_2) L^j(\textbf{k}_2,\textbf{q}_2) \Big. \nonumber \\
&
+
{\cal G}^{\rm NE}_2(k_1^-,k_2^-; L^+)
 \bigg\lgroup (2\pi)^2 \delta^{(2)}(\textbf{k}_1-\textbf{k}_2) + \frac{1}{2} (2\pi)^2 \delta^{(2)}\big[ \textbf{k}_1-\textbf{q}_1-(-\textbf{k}_2+\textbf{q}_2)\big]\bigg\rgroup \nonumber \\
&\hskip 3cm \Big. \times L^i(\textbf{k}_1,\textbf{q}_1) L^i(\textbf{k}_1,\textbf{q}_2) \,  L^j(\textbf{k}_2,\textbf{q}_1) L^j(\textbf{k}_2,\textbf{q}_2) \bigg\}+(\underline{k}_2\rightarrow -\underline{k}_2),
\end{align}
where we have defined the transverse area of the projectile through $(2 \pi)^2 \delta^{(2)}(\textbf{q}-\textbf{q})\to S_\perp$.

Using \cref{eqn9,eqn10,eqn11} we organize the contributions to the double inclusive gluon production cross section, \cref{eqn1},  and finally write it in the following form:
\begin{align}\label{eqn12}
\frac{d \sigma}{d^2 k_1 d\eta_1 d^2k_2 d\eta_2}&=2 (4\pi)^2  \, \alpha_s^2 g^4 \, C_A^2 (N_c^2-1) \, {\cal G}_1^{\rm NE}(k_1^-; \lambda^+) \, {\cal G}_2^{\rm NE}(k_2^-; \lambda^+) \, \mu^4\,  S_\perp 
\nonumber \\
&\times 
\Big\{I_{\rm uncor} +I_{\rm TBE}+I_{\rm PBE,a}+ I_{\rm HBT} +I_{\rm PBE,b}  \Big\}.
\end{align}
The expressions $I_{\rm uncor}$, $I_{\rm TBE}$, $I_{\rm PBE,a}$, $I_{\rm HBT}$ and $I_{\rm PBE,b}$, corresponding to uncorrelated production, Bose enhancement in the target wave function, first piece of Bose enhancement in the projectile wave function, HBT and second piece of Bose enhancement in the projectile wave function respectively, can be found in \ref{app}.
The aim of the next Section will be the analysis of the azimuthal structures in two particle correlations through the standard expansion in Fourier harmonics.

\section{Azimuthal harmonics}
\label{sec:azimuth}

The resulting \cref{eqn12} can be expanded in a Fourier series. Being an even function, only the cosine terms of the series will contribute. That is, we can write
\begin{align}
\frac{d \sigma}{d^2 \textbf{k}_1 d\eta_1 d^2\textbf{k}_2 d\eta_2} \equiv N(k_1,k_2,\Delta \phi)=a_0(k_1,k_2)+\sum_{n=1}^{\infty} a_n(k_1,k_2) \cos(n \Delta \phi),
\end{align}
where $\Delta \phi=\phi_1-\phi_2$ and
\begin{align}
a_n(k_1,k_2)=\frac{2}{\pi(1+\delta_{n0})} \int_{0}^{\pi}  N(k_1,k_2,\Delta \phi) \cos(n \Delta \phi) d\Delta \phi.
\end{align}

We standardly rewrite these series as 
\begin{align}
 N(k_1,k_2,\Delta \phi)=a_0(k_1,k_2)\left[1+\sum_{n=1}^{\infty} 2V_{n\Delta}(k_1,k_2) \cos(n \Delta \phi)\right],
\end{align}
where
\begin{align}
2V_{n\Delta}(k_1,k_2)=\frac{a_n(k_1,k_2)}{a_0(k_1,k_2)}=2 \ \frac{\int_{0}^{\pi}  N(k_1,k_2,\Delta \phi) \cos(n \Delta \phi) d\Delta \phi}{\int_{0}^{\pi}  N(k_1,k_2,\Delta \phi) d\Delta \phi}.
\end{align}

We are interested in studying the dependence of these coefficients on the transverse momentum of one of the produced particles. However, there is some freedom in the definition of this transverse momentum:
\begin{enumerate}[{(}i{)}]
	\item \textit{$k_1=p_T^{ref}$ and $k_2=p_T$}:
	
	One way of defining the $p_T$ dependence of the Fourier coefficients is by fixing one of the momenta, say $k_1$, to a some reference momentum $p_T^{ref}$ and letting the other momentum as the free variable, that is, $k_2=p_T$. With this choice, the azimuthal harmonics are defined as (see e.g. \cite{Chatrchyan:2013nka})	
	\begin{align}\label{k1fix}
	v_n(p_T)=\frac{V_{n\Delta}(p_T,p_T^{ref})}{\sqrt{V_{n\Delta}(p_T^{ref},p_T^{ref})}}.
	\end{align}

	\item \textit{$k_1=k_2=p_T$}:
	
	Another way of fixing the $p_T$ dependence is by setting $k_1=k_2=p_T^{ref}=p_T$ and therefore
	
	\begin{align}\label{k2ptref}
	v_n(p_T)=\sqrt{V_{n\Delta}(p_T,p_T)}.
	\end{align}
	
	\item \textit{Integrating over $k_1$ and $k_2=p_T$}:
	
	Following we can define $v_n(p_T)$ by integrating over $k_1$ and letting $k_2$ free as in \cite{Davy:2018hsl}, that is,
	\begin{align}\label{vnint}
	2v_n^2(p_T)=\frac{\int_{0}^{\infty}k_1 dk_1 a_n(k_1,p_T)}{\int_{0}^{\infty}k_1 dk_1 a_0(k_1,p_T)}=2 \frac{\int_{0}^{\infty}k_1 dk_1\int_{0}^{\pi}  N(k_1,p_T,\Delta \phi) \cos(n \Delta \phi) d\Delta \phi}{\int_{0}^{\infty}k_1 dk_1\int_{0}^{\pi}  N(k_1,p_T,\Delta \phi) d\Delta \phi}\, .
	\end{align}
\end{enumerate}

In the next subsection we will explore the three possibilities.

\subsection{Numerical results}
In order to compute the azimuthal harmonics we first write the non-eikonal correction \cref{eqn6} as
\begin{align}\label{neikon}
{\cal G}^{\rm NE}_2(k_1^-,k_2^-;L^+)=\left\{\frac{\sqrt{2} }{\left(k_1 e^{-\eta_1}-k_2 e^{-\eta_2}\right)L^+} \sin \left[ \frac{\big( k_1 e^{-\eta_1}-k_2 e^{-\eta_2}\big)}{\sqrt{2}} L^+ \right] \right\}^2,
\end{align}
where $\eta_{1,2}$ are the pseudorapidities of the gluons and we use the fact that $k^-=\frac{k^2}{2 k^+}$, $k^+=\frac{1}{\sqrt{2}} k e^{\eta}$.

If $L$ is the size of the target in its rest frame, then we have that $L^+=\frac{1}{\gamma \sqrt{2}}L\approx 2 A^{1/3}/\gamma\ \text{fm}\approx 10A^{1/3}/\gamma \ \text{GeV}^{-1}$, where $A$ is the mass number of the nucleus and $\gamma \simeq \sqrt{s_{\rm NN}}/(2 m_N)$ accounts for the Lorentz contraction in the center of mass frame (therefore, our pseudorapidities will be considered in this frame). Furthermore, for the numerics we take the gluonic size of the projectile to be $B_p = 4 \ \text{GeV}^{-2}$ \cite{Kowalski:2006hc}, $S_\perp=2 \pi B_p \approx 9.8$ mb, $L=12$ fm (Pb nucleus) unless otherwise stated\footnote{As our aim is not to describe experimental data but to discuss the effect of the considered non-eikonal corrections, we will apply the calculation for proton-nucleus collisions though, as indicated above, the Glasma graph approach is only valid for collisions between dilute objects. See \cite{Lappi:2015vta} for a comparison of the results of the Glasma graph approximation with a full dilute-dense numerical computation.} and $N_c=3$. We also take $\lambda^+=0$ in \cref{eq:c1} -- note that this factor is irrelevant for the azimuthal harmonics using definitions in \cref{k1fix} and \cref{k2ptref} and gives a very small contribution using \cref{vnint}.

\begin{figure}
	\centering
	\includegraphics[width=0.9\linewidth]{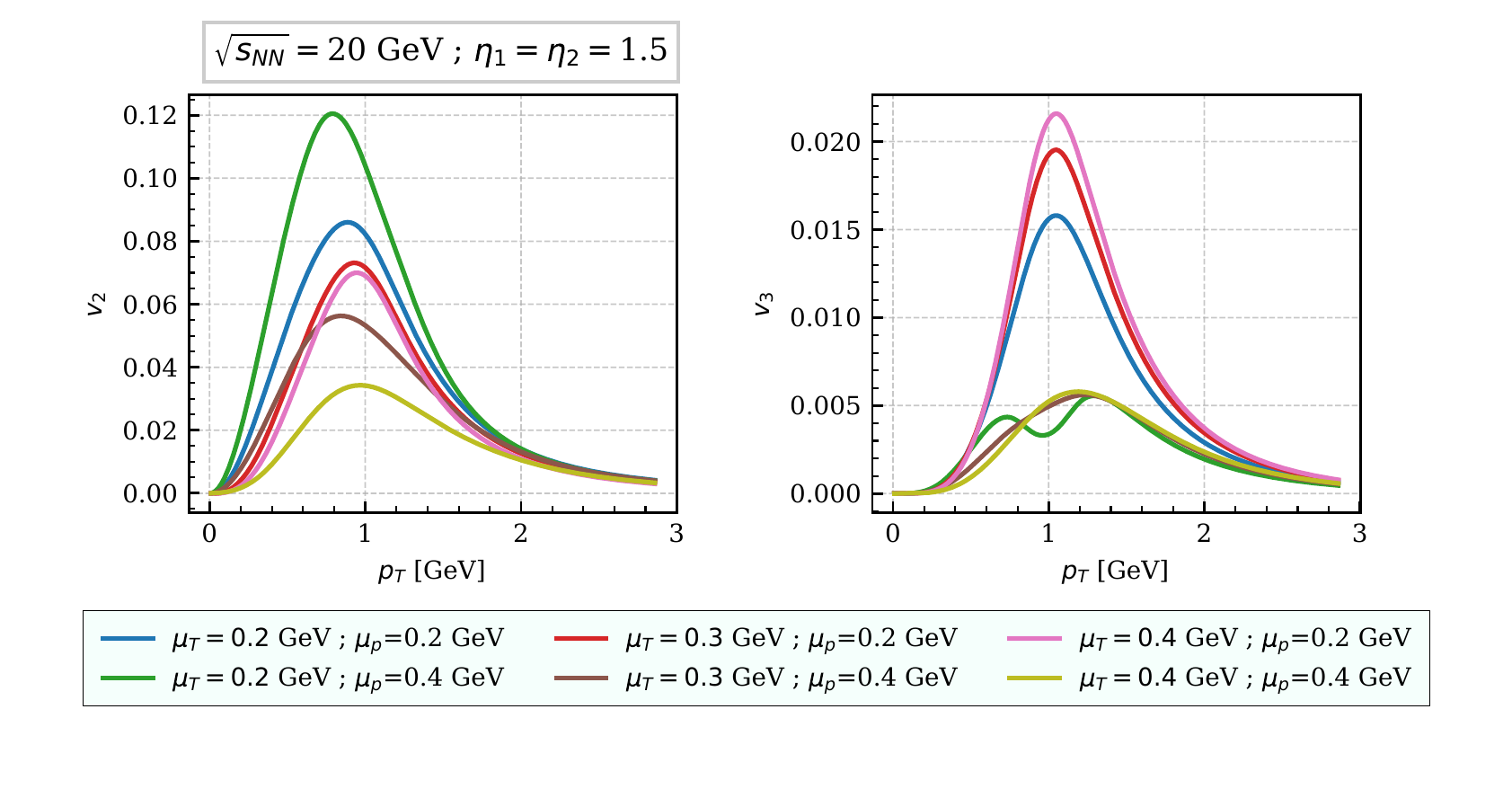}
	\caption{Azimuthal harmonics $v_2(p_T)$ and $v_3(p_T)$ using the definition \cref{k1fix} for different values of $\mu_T$ and $\mu_P$, excluding the HBT contribution, see the text. For these graphs we have taken $p_T^{ref}=1$ GeV, $\sqrt{s_{\rm NN}}=20$ GeV and $\eta_1=\eta_2=1.5$.}
	\label{fig2}
\end{figure}

Taking these values we start by computing the azimuthal harmonics using the definition \cref{k1fix} without the HBT contribution \cref{hbt}. 
It is evident from \cref{eq:yuk} that the Yukawa-type potential for the color fields inside the target has a strong cut-off dependence in the infrared which is analogous to the Debye mass. Since this infrared regulator is related with the Yukawa-type potential introduced to model the target fields, it is denoted by $\mu_T$. Moreover, the product of the Lipatov vertices which defines the emission of the gluons have also cut-off dependence in order to remove the infrared divergences (see \ref{app} for the explicit expressions). This cut-off is denoted as $\mu_P$. Therefore, azimuthal harmonics $v_n(p_T)$ which are calculated  using the Yukawa-type potential and the product of Lipatov vertices, carry a dependence on these two infrared cut-offs $\mu_T$ and $\mu_P$.
We first compute $v_2$ and $v_3$ taking several values of these parameters $\sim \Lambda_{QCD} \approx 0.2 \ \text{GeV}$ in order to see how strong the dependence is. The results are shown in \cref{fig2}. We can see that the height of the peak in $v_2$ becomes smaller as $\mu_T$ gets larger and that the shape is slightly different when $\mu_T \ne \mu_P$, with even a two-peak structure appearing in some case. On the other hand the height of the peak in $v_3$ gets smaller when $\mu_P$ has smaller values. Since in this paper we are interested in the behaviour of the odd azimuthal harmonics, we will use in the rest of the document the values of $\mu_T$ and $\mu_P$ that maximize $v_3$, that is, $\mu_T=0.4 \ \text{GeV}$ and $\mu_P=0.2 \ \text{GeV}$~\footnote{While these values lie close to $\Lambda_{\rm QCD}$, it is difficult to say how realistic they can be considered.}.  As mentioned, we are omitting in these plots a peak around $p_T=p_T^{ref}$ which comes from the HBT contribution.

\begin{figure}[h!]
	\centering
	\includegraphics[width=0.7\linewidth]{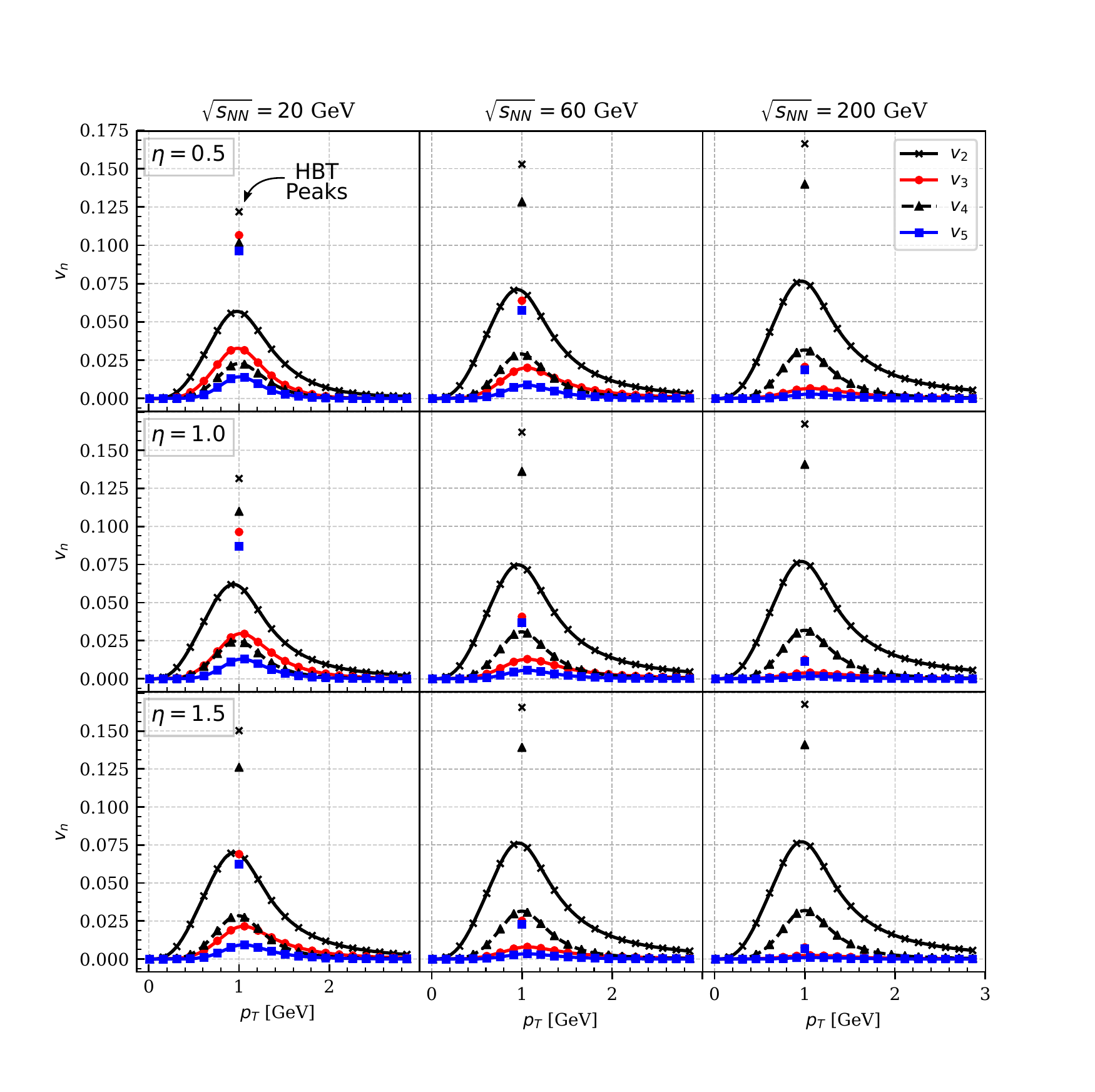}
	\caption{Two particle azimuthal harmonics generated in the non-eikonal Glasma graph approximation, using the definition \cref{k1fix}. The values were calculated using $\mu_T=0.4 \ \text{GeV}$, $\mu_P=0.2 \ \text{GeV}$ and $p_T^{ref}= 1$ GeV at different center of mass energies and gluon pseudorapidities $\eta_1=\eta_2=\eta$. The symbols without lines indicate the HBT contributions.}
	\label{fig4}
\end{figure}

In \cref{fig4} we compute the azimuthal harmonics up to $v_5$ using the definition \cref{k1fix} and taking $p_T^{ref}= 1$ GeV, for different values of $\sqrt{s_{\rm NN}}$ and $\eta_1=\eta_2=\eta$ (we choose $\Delta \eta=\eta_1-\eta_2=0$ in order to maximize the values of the odd harmonics). The HBT contribution coming from \cref{hbt} is also included in this plot. We can see that increasing the center of mass energy decreases the value of the odd harmonics. This is the behaviour that one should expect since when the Lorentz gamma grows up the non-eikonal corrections become smaller. We obtain the same behaviour when increasing the pseudorapidity of the produced gluons. From this plot we conclude that non-eikonal corrections are not important for collisions at high center-of-mass energies as the ones at the LHC but they can be important for collisions at RHIC where $\sqrt{s_{\rm NN}}\le 200$ GeV. However, we note that one should be careful since by going to smaller values of $\sqrt{s_{\rm NN}}$ we are leaving the region of small $x$ or high energies where our formalism can be safely applied.

We would like to also mention that in \cref{fig4}, it is apparent that both even and odd harmonics peak around $p_T\sim p_T^{ref}$. This is due to the fact that $p_T^{ref}$ is chosen to be $k_1$ and $p_T$ is defined as $k_2$, and the values of both even and odd harmonics maximise when the momenta of both produced gluons ($k_1$ and $k_2$ in our notation) are close to each other.

On the other hand, the unrealistic peaked shape of the HBT contribution is due to the fact we have used a simplistic approach, $\mu^2(\textbf{k},\textbf{q})\propto \delta^{(2)}(\textbf{k}+\textbf{q})$. A more realistic approach would employ some function $F[(\textbf{k}+\textbf{q}) \sqrt{B_p}]$ (with $B_p$ being the gluonic size of the projectile), which is peaked around $\textbf{k}+\textbf{q}=0$, e.g. a Gaussian, in which case we should obtain a bell shape with smaller values for the harmonics when $p_T=p_T^{ref}$. We will show results using a Gaussian distribution below.

One interesting behaviour of the odd azimuthal harmonics, stemming from the non-eikonal effects, that we observe in \cref{fig4} is that at any fixed energy the value of the odd harmonics decreases with increasing value of rapidity $\eta$. This behaviour is completely natural since the size of the odd harmonics is directly related the non-eikonal corrections in our framework. When expressed in terms of the rapidity, the eikonal expansion parameter can be written as $p_TL^+e^{-\eta}$. With increasing value of the rapidity, non-eikonal corrections (and therefore odd harmonics) get smaller and vanish completely in the strict eikonal limit.

\begin{figure}[h!]
	\centering
	\includegraphics[width=0.5\linewidth]{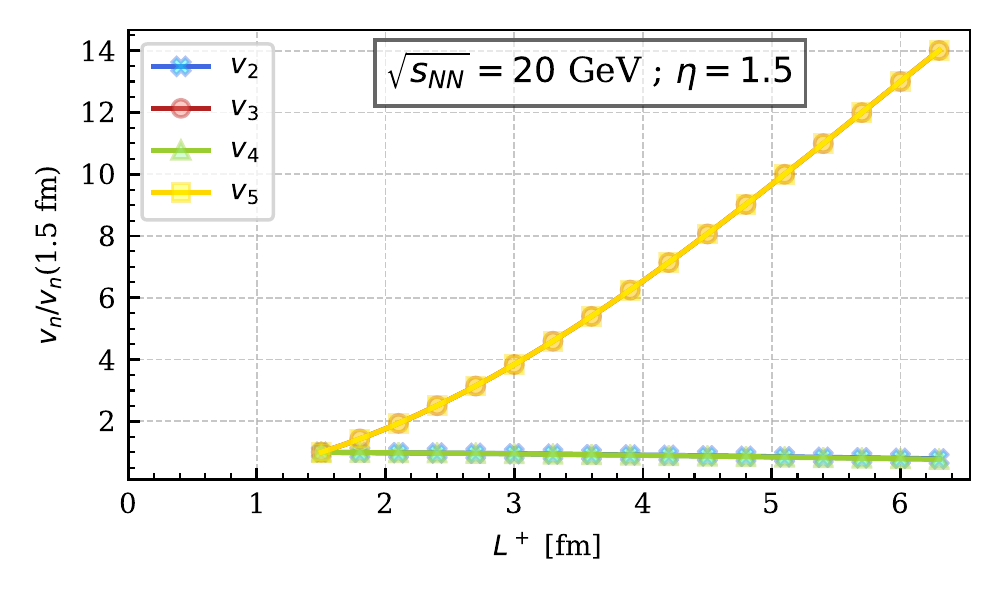}
	\caption{Scaling of $\frac{v_n(L^+)}{v_n(1.5\ {\rm fm})}$ with $L^+$. The odd harmonics increase strongly with increasing $L^+$ while the even ones are almost constant.}
	\label{fig5}
\end{figure}

Another interesting feature of our result is that odd harmonics depend strongly of the size of the target while even ones are almost independent. Furthermore, all odd harmonics and all even harmonics show a good scaling with $L^+$, as can be seen in \cref{fig5}. There we plot $v_n$, using the definition \cref{k1fix}, divided by its value for $L^+=1.5$ fm. While the dependence with centrality and multiplicity would demand a detailed study and the variation of parameters in the model, see e.g. \cite{Mace:2018yvl,Kovner:2018azs}, the increase of $L^+$ with increasing centrality should be one of the ingredients in such dependence and this finding resembles qualitatively that in \cite{Mace:2018yvl}.

\begin{figure}[h!]
	\centering
	\includegraphics[width=0.9\linewidth]{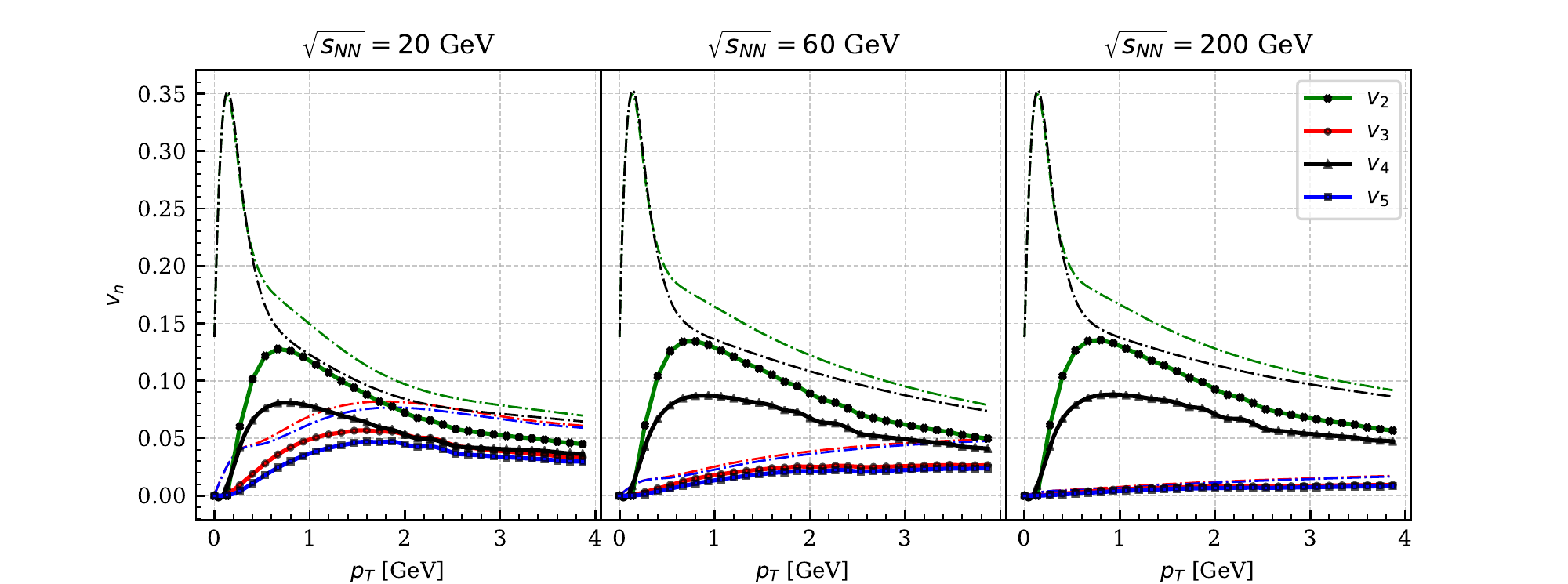}
	\caption{Azimuthal harmonics computed using the prescription of \cref{k2ptref}. The parameters used for this plot are $\mu_T=\mu_g=0.4 \ \text{GeV}$, $\mu_P=0.2 \ \text{GeV}$ and $\eta_1=\eta_2=1.5$. The dashed lines are the result using $\mu^2(\textbf{k}_1,\textbf{k}_2) \propto (2\pi)^2 \delta^{(2)} (\textbf{k}_1-\textbf{k}_2) $ and the continuous lines employ $\mu^2(\textbf{k}_1,\textbf{k}_2) \propto  2 \pi B_p \exp{\big(-\frac{B_p}{2}(\textbf{k}_1-\textbf{k}_2)^2\big)} $.}
	\label{fig6}
\end{figure}

\begin{figure}[h!]
	\centering
	\includegraphics[width=0.9\linewidth]{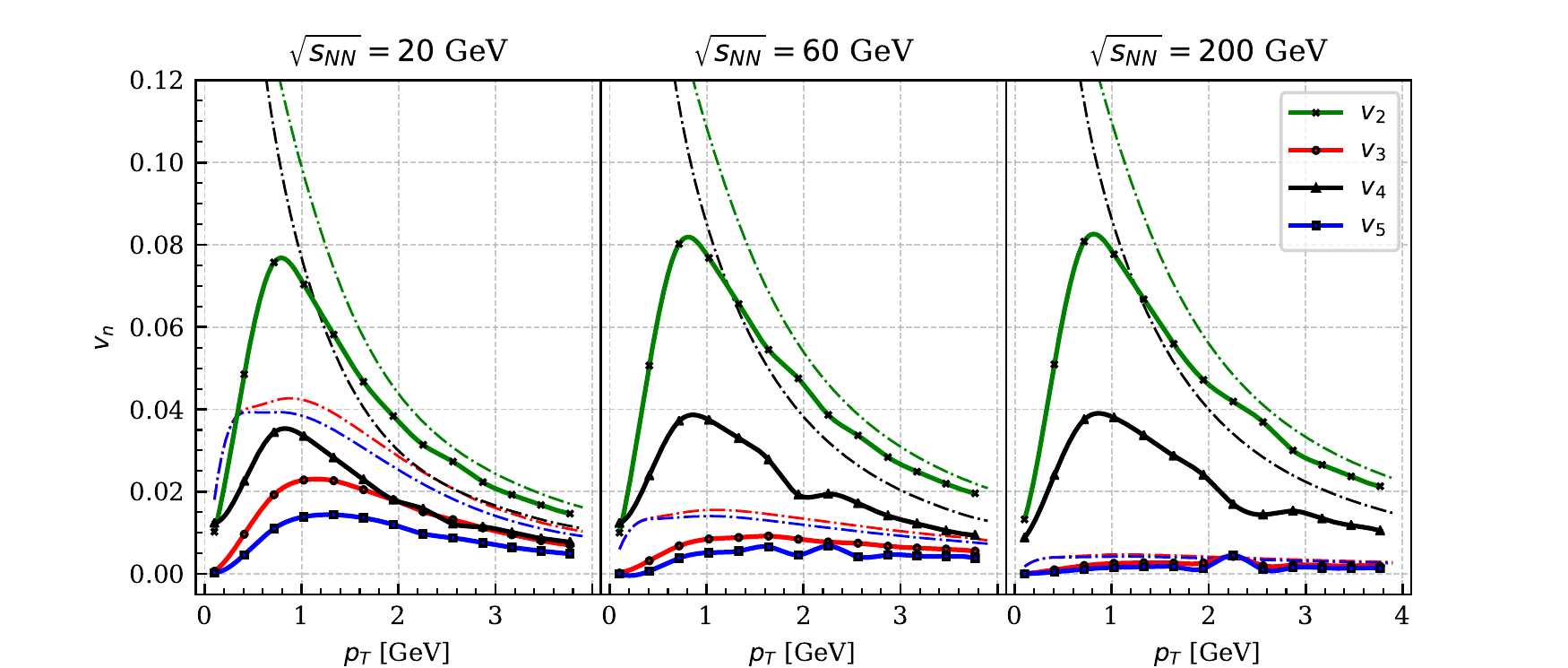}
	\caption{Azimuthal harmonics computed using the prescription of \cref{vnint}. The parameters used for this plot are $\mu_T=\mu_g=0.4 \ \text{GeV}$, $\mu_P=0.2 \ \text{GeV}$ and $\eta_1=\eta_2=1.5$. The dashed lines are the result using $\mu^2(\textbf{k}_1,\textbf{k}_2) \propto (2\pi)^2 \delta^{(2)} (\textbf{k}_1-\textbf{k}_2) $ and the continuous lines employ $\mu^2(\textbf{k}_1,\textbf{k}_2) \propto  2 \pi B_p \exp{\big(-\frac{B_p}{2}(\textbf{k}_1-\textbf{k}_2)^2\big)} $.}
	\label{fig7}
\end{figure}

For the sake of completeness, we also compute the azimuthal harmonics using prescriptions  \cref{k2ptref} and \cref{vnint}. Now, since we are integrating over  variable $k_1$, we have to regulate the $1/k^2$ term that arises in \cref{lipatov}. In order to do so, we just substitute $1/k^2\rightarrow 1/(k^2+\mu_g^2)$ and we choose $\mu_g=0.4$ GeV~\footnote{The effect of changing this value to 0.2 GeV affects the azimuthal harmonics for $p_T< 0.5$ GeV when the delta function form of the HBT term is used. For a Gaussian, see below, no sizeable effect of this change of $\mu_g$ is observed.}. The results are shown in \cref{fig6} and \cref{fig7} where we have used $\mu_T=0.4 \ \text{GeV}$, $\mu_P=0.2 \ \text{GeV}$ and $\eta_1=\eta_2=1.5$. The dashed lines are our results for a Dirac delta in $\mu^2(\textbf{k}_1,\textbf{k}_2)$, and we observe that the shape of $v_n(p_T)$ is very abrupt and unrealistic for small $p_T$. This is what we should expect since $\mu^2(\textbf{k}_1,\textbf{k}_2) \propto (2\pi)^2 \delta^{(2)} (\textbf{k}_1-\textbf{k}_2) $ comes from assuming translational invariance and this is only valid for large $|\textbf{k}_1-\textbf{k}_2|$ or $B_p$ but, in our case, we are using  small values for both $|\textbf{k}_1-\textbf{k}_2|$ and $B_p$. In order to deal with this problem we make the substitution $(2\pi)^2 \delta^{(2)} (\textbf{k}_1-\textbf{k}_2)\rightarrow 2 \pi B_p \exp{\big(-\frac{B_p}{2}(\textbf{k}_1-\textbf{k}_2)^2\big)}$ in the HBT term \cref{hbt} since this is the dominant contribution. The corresponding results can be seen in the continuous lines of \cref{fig6} and \cref{fig7} and they are smoother.

Writing \cref{neikon} as 
\begin{align}
{\cal G}^{\rm NE}_2(k_1^-,k_2^-;L^+)=\left\{ \frac{\sqrt{2}\,  e^{\eta_1} }{\big(k_1-k_2 \, e^{\Delta \eta}\big)L^+}    \sin \left[\frac{\big(k_1-k_2\,  e^{\Delta \eta}\big)}{\sqrt{2}} e^{-\eta_1} L^+ \right]  \right\}^2,
\end{align}
we can study the dependence of the cross section with respect to the difference in rapidity between the produced particles given by the non-eikonal corrections -- this dependence is absent if non-eikonal corrections are neglected.

\begin{figure}
	\centering
	\includegraphics[width=0.5\linewidth]{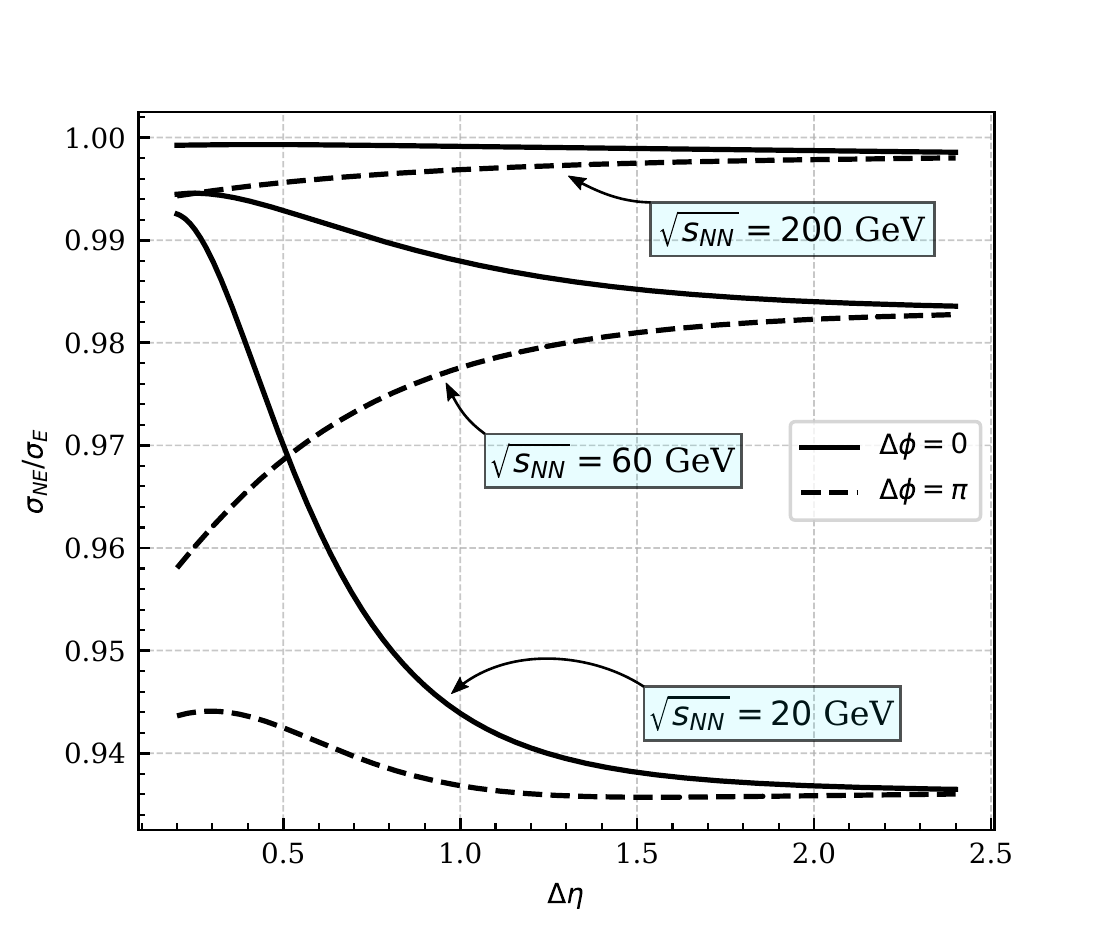}
	\caption{Ratio of non-eikonal cross section with respect to the eikonal one both in the forward ($\Delta \phi=0$) and the backward ($\Delta \phi=\pi$) peaks and for $\eta_1=0$, $k_1=1$ GeV and $k_2=1.2$ GeV.}
	\label{fig9}
\end{figure}

In \cref{fig9} we have plotted the ratio of the non-eikonal cross section with respect to the eikonal one both in the forward ($\Delta \phi=0$) and the backward ($\Delta \phi=\pi$) peaks and for $\eta_1=0$, $k_1=1$ GeV and $k_2=1.2$ GeV (we set $k_1\ne k_2$ to not include the HBT contribution). We can see that there is a sizeable difference between the peaks up to $1.5-2$ units in rapidity for $\sqrt{s_{\rm NN}}=20$ and $60$ GeV, and that the difference becomes negligible for higher energies, $\sqrt{s_{\rm NN}}=200$ GeV, as expected. 

\begin{figure}
	\centering
	\includegraphics[width=0.7\linewidth]{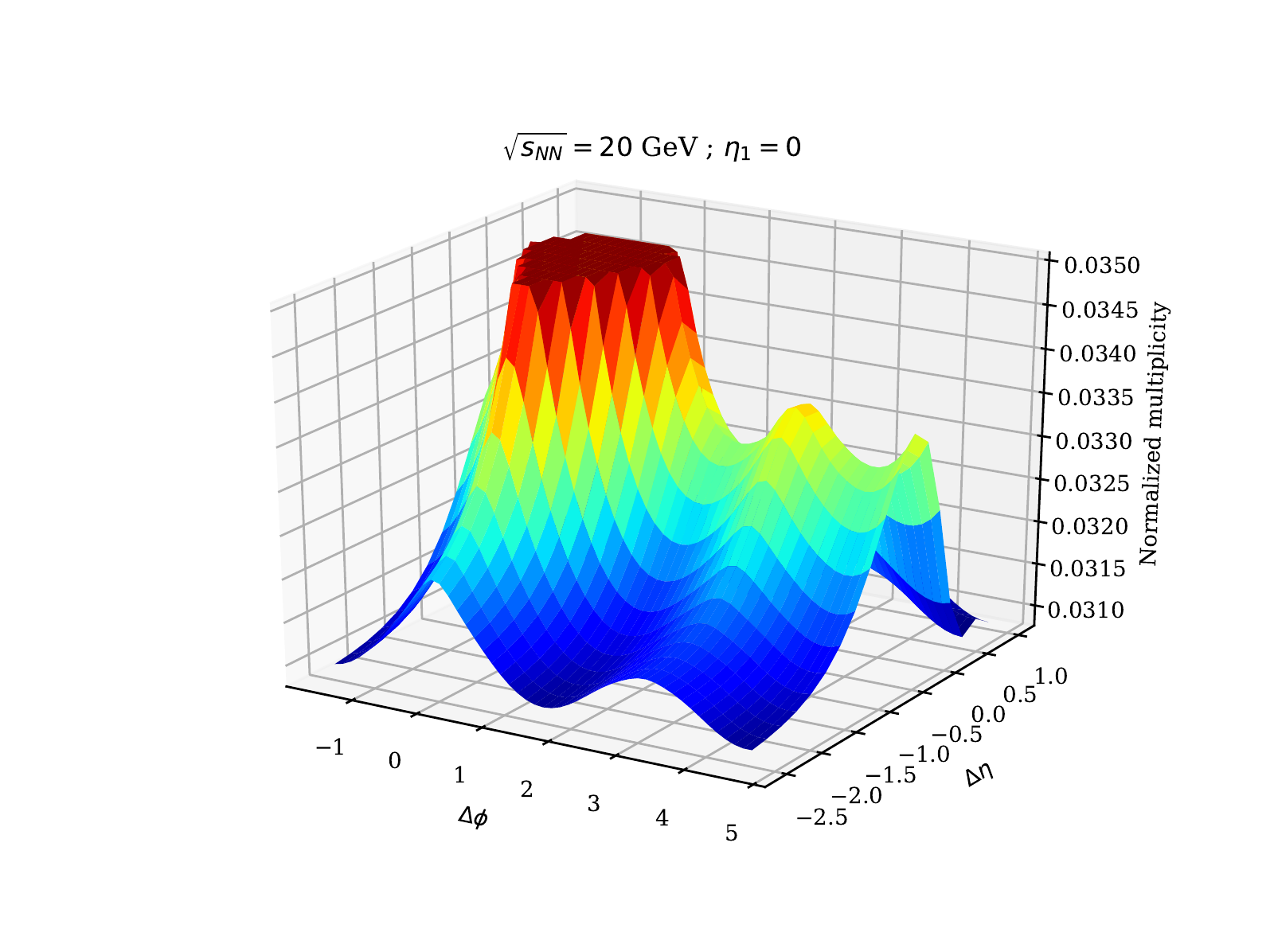}
	\caption{Dependence of the normalized multiplicity with respect to the azimuthal angle and the rapidity difference. In order to make the non-eikonal effects more visible, we have cut the near-side peak for values larger than 0.0345.}
	\label{fig8}
\end{figure}

 In \cref{fig8} we plot the cross section \cref{eqn12} without the prefactors outside the curly brackets (that we call normalized multiplicity) against $\Delta \eta$ and $\Delta \phi$ using $\eta_1=0$, $k_1=1$ GeV and $k_2=1.2$ GeV. We can see again that the differences between the forward and backward peaks are visible up to 2.5 pseudorapidity units.

\section{Conclusions}
\label{sec:conclu}

In this manuscript we have analyzed the effect on the non-eikonal corrections stemming from relaxing the shockwave approximation for the target which, therefore, acquires a finite length, on the two gluon inclusive cross section in the CGC. We work in the Glasma graph approximation suitable for collisions between dilute objects (pp). While the corresponding expressions were derived in a previous publication~\cite{Agostini:2019avp}, here we focus on the numerical implementation, for which several model assumptions are made. We make no attempt to compare with experimental data but only address the existence and size of the non-eikonal effects  on the azimuthal structure.

We explore how the non-eikonal corrections break the accidental forward-backward symmetry present in usual CGC calculations, and thus lead to sizeable odd harmonics. We discuss the different contributions: Bose enhancement of the projectile and target wave functions and HBT, and check the stability of the qualitative behavior of the results against variations in the functional forms and parameters in the model assumptions. We find a good scaling of all even and all odd harmonics with respect to the length of the target, with even harmonics being constant and odd ones growing with increasing length.
The non-eikonal corrections vanish with increasing energy of the collision, being sizeable up to the top energies at RHIC but negligible for those at the LHC. Furthermore, they turn to be significant for pseudorapidity differences between the produced gluons up to about 2.5 units. Therefore, we conclude that non-eikonal effects cannot be the dominant source of odd harmonics at the highest energies but they can be relevant for those at RHIC.

The outlook of this work is its extension to dilute-dense (pA) collisions that will be the subject of a forthcoming publication, and a comparison to experimental data.

\section*{Acknowledgements}
PA and NA are supported by  Ministerio de Ciencia e Innovaci\'on of Spain under projects FPA2017-83814-P and Unidad de Excelencia Mar\'{\i}a de Maetzu under project MDM-2016-0692,  by Xunta de Galicia under project ED431C 2017/07, and by FEDER.
The work of TA is supported by Grant No. 2018/31/D/ST2/00666 (SONATA 14 - National Science Centre, Poland).
This work has been performed in the framework of COST Action CA15213 ``Theory of hot matter and relativistic heavy-ion collisions" (THOR), MSCA RISE 823947  ``Heavy ion collisions: collectivity and precision in saturation
physics" (HIEIC) and   has received funding from the European Union's Horizon 2020 research and innovation programme under grant agreement No 824093.

\appendix
\section{Expressions for $I_{\rm uncor}$, $I_{\rm TBE}$, $I_{\rm PBE,a}$, $I_{\rm HBT}$ and $I_{\rm PBE,b}$}
\label{app}

In this Appendix we write the expressions for the different terms in \cref{eqn12} and put them in a form suitable for numerical computation.

\subsection{The fully uncorrelated term: $I_{\rm uncor}$}

The first term that we will evaluate is $I_{\rm uncor}$. This term corresponds to the fully uncorrelated two gluon production. Although it does not contribute to the azimuthal harmonics, it gives the bulk of the two gluon cross section. Therefore, it is important in order to have it properly normalized.
From \cref{eqn9}, we have that
\begin{align}\label{eqn13}
I_{\rm uncor} = S_{\perp} (N_c^2-1) \int \frac{d^2 \textbf{q}_1}{(2 \pi)^2} \frac{d^2 \textbf{q}_2}{(2 \pi)^2} \frac{\mu_T^2}{(q_1^2+\mu_T^2)^2} \frac{\mu_T^2}{(q_2^2+\mu_T^2)^2} L^i(\textbf{k}_1,\textbf{q}_1) L^i(\textbf{k}_1,\textbf{q}_1) L^j(\textbf{k}_2,\textbf{q}_2) L^j(\textbf{k}_2,\textbf{q}_2),
\end{align}
where the product of two Lipatov vertices is
\begin{align}\label{eqn14}
L^i(\textbf{k},\textbf{q}) L^i(\textbf{k},\textbf{q})=
\frac{q^2}{k^2[k^2+q^2+\mu_P^2-2 k q \cos \phi]}\,,
\end{align}
with $\phi$ is the angle between $\textbf{k}$ and $\textbf{q}$. Furthermore we have introduced an infrared regulator $\mu_P$ in the momentum of the projectile gluon to regulate this otherwise divergent integral\footnote{Were the projectile a dense object, the natural infrared regulator would be its saturation scale.}.

Since  \cref{eqn13} decouples into two identical integrals, the only integral that we have to deal with is
\begin{align}
\int \frac{d^2 \textbf{q}}{(2 \pi)^2} \frac{\mu_T^2}{(q^2+\mu_T^2)^2} L^i(\textbf{k},\textbf{q}) L^i(\textbf{k},\textbf{q})&=\frac{1}{(2 \pi)^2} \int_0^\infty q dq \frac{\mu_T^2}{(q^2+\mu_T^2)^2} \int_{0}^{2 \pi} d\phi \frac{q^2}{k^2[k^2+q^2+\mu_P^2-2 k q \cos \phi]}
\nonumber \\
&=\frac{\mu_T^2}{2 \pi} \int_0^\infty dq \frac{q}{(q^2+\mu_T^2)^2} \frac{q^2}{k^2\sqrt{(k^2+q^2+\mu_P^2)^2-4 k^2 q^2} }\, .
\end{align}
This integral can be solved analytically but the expression is rather lengthy.

\subsection{Bose enhancement in the target wave function: $I_{\rm TBE}$}

The second term that we will have to deal is $I_{\rm TBE}$. This term fixes $\textbf{q}_1=\textbf{q}_2$ and therefore it is a Bose enhancement in the target wave function. From \cref{eqn10}, we can write it as
\begin{align}
I_{\rm TBE}&= {\cal G}^{\rm NE}_2(k_1^-,k_2^-;L^+) \int \frac{d^2 \textbf{q}_1}{(2 \pi)^2} \frac{d^2 \textbf{q}_2}{(2 \pi)^2} \frac{\mu_T^2}{(q_1^2+\mu_T^2)^2} \frac{\mu_T^2}{(q_2^2+\mu_T^2)^2} \big[ (2\pi)^2 \delta^{(2)}(\textbf{q}_1-\textbf{q}_2)\big]
\nonumber\\
&\times
 L^i(\textbf{k}_1,\textbf{q}_1) L^i(\textbf{k}_1,\textbf{q}_2) \, L^j(\textbf{k}_2,\textbf{q}_2) L^j(\textbf{k}_2,\textbf{q}_1) 
+(\underline{k}_2\rightarrow -\underline{k}_2) \nonumber \\
&= {\cal G}_2^{\rm NE}(k_1^-,k_2^-;L^+) \int \frac{d^2 \textbf{q}}{(2 \pi)^2} \frac{\mu_T^4}{(q^2+\mu_T^2)^4} L^i(\textbf{k}_1,\textbf{q}) L^i(\textbf{k}_1,\textbf{q}) L^j(\textbf{k}_2,\textbf{q}) L^j(\textbf{k}_2,\textbf{q}) +(\underline{k}_2\rightarrow -\underline{k}_2).
\end{align}

 Using \cref{eqn14} and
 \begin{align}
 \int_{0}^{2 \pi} d\phi \frac{1}{a+\cos \phi}\frac{1}{b+\cos (\phi-\Delta \phi)}=2\pi \frac{\frac{a}{\sqrt{a^2-1}}+\frac{b}{\sqrt{b^2-1}}}{a b +\sqrt{a^2-1}\sqrt{b^2-1} -\cos \Delta \phi}
 \end{align}
 to solve the $\phi$ integral, we obtain
\begin{align}
I_{\rm TBE}&= \frac{\mu_T^4}{2 \pi k_1^2 k_2 ^2} \, {\cal G}^{\rm NE}_2(k_1^-,k_2^-; L^+) \int_{0}^{\infty} dq  \frac{q^5}{\left(\mu_T^2+q^2\right)^4} \nonumber \\
&\times \frac{\frac{1}{\sqrt{1-\frac{4 k_1^2 q^2}{\left(k_1^2+\mu_P^2+q^2\right)^2}}}+\frac{1}{\sqrt{1-\frac{4 k_2^2 q^2}{\left(k_2^2+\mu_P^2+q^2\right)^2}}}}{\left(k_1^2+\mu_P^2+q^2\right) \left(k_2^2+\mu_P^2+q^2\right)+\sqrt{\left(k_1^2+\mu_P^2+q^2\right)^2-4 k_1^2 q^2}
	\sqrt{\left(k_2^2+\mu_P^2+q^2\right)^2-4 k_2^2 q^2}-4 k_1 k_2 q^2 \cos \Delta \phi } \nonumber \\
&+[(k_2^-,\Delta \phi)  \rightarrow (-k_2^-,\Delta \phi + \pi) ],
\end{align}
where $\Delta \phi$ is the angle between $\textbf{k}_1$ and $\textbf{k}_2$ -- the azimuthal correlation angle that we are interested to study.

\subsection{Bose enhancement in the projectile wave function: $I_{\rm PBE,a}$}

We will now separate the quadrupole contribution, \cref{eqn11}, into tree parts. The first of them,  $I_{\rm PBE,a}$, fixes $\textbf{k}_1-\textbf{q}_1=\textbf{k}_2-\textbf{q}_2$ and therefore it is a Bose enhancement in the projectile wave function. This term does not contain any non-eikonal correction and therefore it has the accidental symmetry $(\textbf{{k}}_2\rightarrow -\textbf{{k}}_2)$ which implies, see section \ref{sec:azimuth}, that it does not generate odd azimuthal harmonics. From the first term in \cref{eqn11} we have 
\begin{align}
I_{\rm PBE,a}&=\int \frac{d^2 \textbf{q}_1}{(2 \pi)^2} \frac{d^2 \textbf{q}_2}{(2 \pi)^2} \frac{\mu_T^2}{(q_1^2+\mu_T^2)^2} \frac{\mu_T^2}{(q_2^2+\mu_T^2)^2} 
\, 
\Big\lgroup (2\pi)^2 \delta^{(2)}\big[\textbf{k}_1-\textbf{q}_1-(\textbf{k}_2-\textbf{q}_2)\big]\Big\rgroup \nonumber\\
&
\hskip 0.5cm
\times 
 L^i(\textbf{k}_1,\textbf{q}_1) L^i(\textbf{k}_1,\textbf{q}_1) \,  L^j(\textbf{k}_2,\textbf{q}_2) L^j(\textbf{k}_2,\textbf{q}_2)
+(\underline{k}_2\rightarrow -\underline{k}_2) \nonumber \\
&=\int \frac{d^2 \textbf{q}}{(2 \pi)^2} \frac{\mu_T^2}{(q^2+\mu_T^2)^2} \frac{\mu_T^2}{\big[ (\textbf{q}+\textbf{k}_2-\textbf{k}_1)^2+\mu_T^2\big]^2}
\nonumber\\
&
\hskip 0.5cm
\times
 L^i(\textbf{k}_1,\textbf{q}) L^i(\textbf{k}_1,\textbf{q}) L^j(\textbf{k}_2,\textbf{q}+\textbf{k}_2-\textbf{k}_1) L^j(\textbf{k}_2,\textbf{q}+\textbf{k}_2-\textbf{k}_1) 
+(\underline{k}_2\rightarrow -\underline{k}_2).\end{align}

This expression can be written  in terms of modulus and angles of vectors using \cref{eqn14}, 
\begin{align}
L^j(\textbf{k}_2,\textbf{q}+\textbf{k}_2-\textbf{k}_1) L^j(\textbf{k}_2,\textbf{q}+\textbf{k}_2-\textbf{k}_1)=\frac{-2 k_1 k_2 \cos (\Delta \phi )+k_2^2+2 k_2 q \cos (\phi - \Delta \phi)}{k_2^2 \left(k_1^2-2 k_1
	q \cos (\phi )+\mu_P^2+q^2\right)}+\frac{1}{k_2^2}
\end{align}
and 
\begin{align}
\label{eq:yukmanip}
\frac{\mu_T^2}{\big[(\textbf{q}+\textbf{k}_2-\textbf{k}_1)^2+\mu_T^2\big]^2}=\frac{\mu_T^2}
{\big[ \mu_T^2+q^2+k_1^2+k_2^2-2 k_1 k_2 \cos (\Delta \phi )-2 k_1 q \cos (\phi )+2 k_2	q \cos (\Delta \phi -\phi )\big]^2}.
\end{align}
While the $\phi$ integral can be solved using the residue theorem, the result is very lengthy and we have decided to deal with both the $\phi$ and  $q$ integrals numerically.

\subsection{HBT contribution: $I_{\rm HBT}$}

The second part of the quadrupole term, $I_{\rm HBT}$, fixes $\textbf{k}_1=\textbf{k}_2$ and therefore it is an HBT contribution. From the second term in \cref{eqn11} we get
\begin{align}\label{hbt}
I_{\rm HBT}&=\int \frac{d^2 \textbf{q}_1}{(2 \pi)^2} \frac{d^2 \textbf{q}_2}{(2 \pi)^2} \frac{\mu_T^2}{({q}_1^2+\mu_T^2)^2} \frac{\mu_T^2}{({q}_2^2+\mu_T^2)^2} \, 
{\cal G}_2^{\rm NE}(k_1^-,k_2^-; L^+) \, 
\big[ (2\pi)^2 \delta^{(2)}(\textbf{k}_1-\textbf{k}_2) \big] \nonumber\\
& \hskip 0.5cm
\times 
L^i(\textbf{k}_1,\textbf{q}_1) L^i(\textbf{k}_1,\textbf{q}_2) \,  L^j(\textbf{k}_2,\textbf{q}_1) L^j(\textbf{k}_2,\textbf{q}_2) 
+(\underline{k}_2\rightarrow -\underline{k}_2) \nonumber \\
&={\cal G}_2^{\rm NE}(k_1^-,k_2^-;L^+) \, \big[ (2\pi)^2 \delta^{(2)}(\textbf{k}_1-\textbf{k}_2)\big] \,  \text{Tr}[M \cdot M] +(\underline{k}_2\rightarrow -\underline{k}_2) \nonumber \\
&=\bigg[ (2\pi)^2 \frac{1}{k_1}\delta({k}_1-{k}_2)\bigg] \,  \text{Tr}[M \cdot M] \Big\{ {\cal G}^{\rm NE}_2(k_1^-,k_2^-; L^+) \delta(\Delta \phi) + {\cal G}^{\rm NE}_2(k_1^-,-k_2^-;L^+) \delta(\Delta \phi - \pi) \Big\},
\end{align}
where we have used that $\delta^{(2)}(\textbf{k}_1-\textbf{k}_2)=\frac{1}{k_1}\delta({k}_1-{k}_2)\delta(\Delta \phi)$ and defined
\begin{align}
M^{ij}=\int \frac{d^2 \textbf{q}}{(2 \pi)^2} \frac{\mu_T^2}{({q}^2+\mu_T^2)^2} L^i(\textbf{k}_1,\textbf{q}) L^j(\textbf{k}_1,\textbf{q}).
\end{align}
The matrix $M$ can be easily solved by taking into account the definition of the Lipatov vertex.

\subsection{Bose enhancement in the projectile wave function: $I_{\rm PBE,b}$}

The last part of the quadrupole term,  $I_{\rm PBE,b}$, fixes $\textbf{k}_1-\textbf{q}_1=-\textbf{k}_2+\textbf{q}_2$ and therefore it is a Bose enhancement term in the projectile wave function. From the third term in \cref{eqn11} we have
\begin{align}
I_{\rm PBE,b}&=\frac{1}{2} \, {\cal G}^{\rm NE}_2(k_1^-,k_2^-;L^+) \int \frac{d^2 \textbf{q}_1}{(2 \pi)^2} \frac{d^2 \textbf{q}_2}{(2 \pi)^2} \frac{\mu_T^2}{(q_1^2+\mu_T^2)^2} \frac{\mu_T^2}{(q_2^2+\mu_T^2)^2}  \, 
\Big\lgroup (2\pi)^2 \delta^{(2)}\big[ \textbf{k}_1-\textbf{q}_1-(-\textbf{k}_2+\textbf{q}_2)\big]\Big\rgroup \nonumber\\
& \hskip 0.5cm
\times
 L^i(\textbf{k}_1,\textbf{q}_1) L^i(\textbf{k}_1,\textbf{q}_2)  \, 
L^j(\textbf{k}_2,\textbf{q}_1) L^j(\textbf{k}_2,\textbf{q}_2)+(\underline{k}_2\rightarrow -\underline{k}_2) \nonumber \\
&=\frac{1}{2}\, {\cal G}_2^{\rm NE}(k_1^-,k_2^-;L^+)  \int \frac{d^2 \textbf{q}}{(2 \pi)^2} \frac{\mu_T^2}{(q^2+\mu_T^2)^2} \frac{\mu_T^2}{\big[(-\textbf{q}+\textbf{k}_1+\textbf{k}_2)^2+\mu_T^2\big]^2} \nonumber\\
&\hskip 0.5cm
\times
 L^i(\textbf{k}_1,\textbf{q}) L^i(\textbf{k}_1,-\textbf{q}+\textbf{k}_1+\textbf{k}_2)  \, 
 L^j(\textbf{k}_2,\textbf{q}) L^j(\textbf{k}_2,-\textbf{q}+\textbf{k}_1+\textbf{k}_2)+(\underline{k}_2\rightarrow -\underline{k}_2).
\end{align}

In order to write this term just in terms of modulus and angles we have to have into account that
\begin{align}
&L^i(\textbf{k}_1,\textbf{q}) L^i(\textbf{k}_1,-\textbf{q}+\textbf{k}_1+\textbf{k}_2)=-\frac{k_1^2-k_1 q \cos (\phi )}{k_1^2 
\big[k_1^2-2 k_1 q \cos (\phi )+\mu_P^2+q^2\big]}+\frac{1}{k_1^2} \\
&-\frac{k_1 q \cos (\phi )-k_1 k_2 \cos (\Delta \phi )}{k_1^2 \big[k_2^2-2 k_2 q \cos (\Delta \phi -\phi )+\mu_P^2+q^2\big]}
+
\frac{-k_1 k_2 \cos (\Delta \phi )+k_1 q \cos (\phi )+k_2 q \cos (\Delta \phi -\phi)-q^2}{\big[k_1^2-2 k_1 q \cos (\phi )+\mu_P^2+q^2\big] \,  \big[k_2^2-2 k_2 q \cos (\Delta \phi -\phi )+\mu_P^2+q^2\big]} \nonumber
\end{align}
and
\begin{align}
&L^j(\textbf{k}_2,\textbf{q}) L^j(\textbf{k}_2,-\textbf{q}+\textbf{k}_1+\textbf{k}_2)=-\frac{k_2^2-k_2 q \cos (\Delta \phi -\phi )}{k_2^2 \big[k_2^2-2 k_2 q \cos (\Delta \phi -\phi )+\mu_P^2+q^2\big]}+\frac{1}{k_2^2} \nonumber \\
&+\frac{-k_1 k_2 \cos (\Delta \phi )+k_1 q \cos (\phi )+k_2 q \cos (\Delta \phi -\phi )-q^2}{\big[k_1^2-2 k_1 q \cos(\phi )+\mu_P^2+q^2\big] \big[k_2^2-2 k_2 q \cos (\Delta \phi -\phi )+\mu_P^2+q^2\big]}-\frac{k_2 q \cos(\Delta \phi -\phi )-k_1 k_2 \cos (\Delta \phi )}{k_2^2 \big[ k_1^2-2 k_1 q \cos (\phi )+\mu_P^2+q^2\big]}.
\end{align}
The expression for $|a(-\textbf{q}+\textbf{k}_1+\textbf{k}_2)|^2$ is similar  to \cref{eq:yukmanip}.

\vskip 1cm

\begin{thebibliography}{78}
\expandafter\ifx\csname natexlab\endcsname\relax\def\natexlab#1{#1}\fi
\providecommand{\url}[1]{\texttt{#1}}
\providecommand{\href}[2]{#2}
\providecommand{\path}[1]{#1}
\providecommand{\DOIprefix}{doi:}
\providecommand{\ArXivprefix}{arXiv:}
\providecommand{\URLprefix}{URL: }
\providecommand{\Pubmedprefix}{pmid:}
\providecommand{\doi}[1]{\href{http://dx.doi.org/#1}{\path{#1}}}
\providecommand{\Pubmed}[1]{\href{pmid:#1}{\path{#1}}}
\providecommand{\bibinfo}[2]{#2}
\ifx\xfnm\relax \def\xfnm[#1]{\unskip,\space#1}\fi
\bibitem[{Khachatryan et~al.(2010)}]{Khachatryan:2010gv}
\bibinfo{author}{V.~Khachatryan}, et~al. (\bibinfo{collaboration}{CMS}),
\newblock \bibinfo{title}{{Observation of Long-Range Near-Side Angular
  Correlations in Proton-Proton Collisions at the LHC}},
\newblock \bibinfo{journal}{JHEP} \bibinfo{volume}{09} (\bibinfo{year}{2010})
  \bibinfo{pages}{091}. \DOIprefix\doi{10.1007/JHEP09(2010)091}.
  \href{http://arxiv.org/abs/1009.4122}{{\tt arXiv:1009.4122}}.
\bibitem[{Khachatryan et~al.(2016)}]{Khachatryan:2015lva}
\bibinfo{author}{V.~Khachatryan}, et~al. (\bibinfo{collaboration}{CMS}),
\newblock \bibinfo{title}{{Measurement of long-range near-side two-particle
  angular correlations in pp collisions at $\sqrt s =$13 TeV}},
\newblock \bibinfo{journal}{Phys. Rev. Lett.} \bibinfo{volume}{116}
  (\bibinfo{year}{2016}) \bibinfo{pages}{172302}.
  \DOIprefix\doi{10.1103/PhysRevLett.116.172302}.
  \href{http://arxiv.org/abs/1510.03068}{{\tt arXiv:1510.03068}}.
\bibitem[{Aad et~al.(2016)}]{Aad:2015gqa}
\bibinfo{author}{G.~Aad}, et~al. (\bibinfo{collaboration}{ATLAS}),
\newblock \bibinfo{title}{{Observation of Long-Range Elliptic Azimuthal
  Anisotropies in $\sqrt{s}=$13 and 2.76 TeV $pp$ Collisions with the ATLAS
  Detector}},
\newblock \bibinfo{journal}{Phys. Rev. Lett.} \bibinfo{volume}{116}
  (\bibinfo{year}{2016}) \bibinfo{pages}{172301}.
  \DOIprefix\doi{10.1103/PhysRevLett.116.172301}.
  \href{http://arxiv.org/abs/1509.04776}{{\tt arXiv:1509.04776}}.
\bibitem[{Chatrchyan et~al.(2013)}]{CMS:2012qk}
\bibinfo{author}{S.~Chatrchyan}, et~al. (\bibinfo{collaboration}{CMS}),
\newblock \bibinfo{title}{{Observation of long-range near-side angular
  correlations in proton-lead collisions at the LHC}},
\newblock \bibinfo{journal}{Phys. Lett.} \bibinfo{volume}{B718}
  (\bibinfo{year}{2013}) \bibinfo{pages}{795--814}.
  \DOIprefix\doi{10.1016/j.physletb.2012.11.025}.
  \href{http://arxiv.org/abs/1210.5482}{{\tt arXiv:1210.5482}}.
\bibitem[{Abelev et~al.(2013)}]{Abelev:2012ola}
\bibinfo{author}{B.~Abelev}, et~al. (\bibinfo{collaboration}{ALICE}),
\newblock \bibinfo{title}{{Long-range angular correlations on the near and away
  side in $p$-Pb collisions at $\sqrt{s_{NN}}=5.02$ TeV}},
\newblock \bibinfo{journal}{Phys. Lett.} \bibinfo{volume}{B719}
  (\bibinfo{year}{2013}) \bibinfo{pages}{29--41}.
  \DOIprefix\doi{10.1016/j.physletb.2013.01.012}.
  \href{http://arxiv.org/abs/1212.2001}{{\tt arXiv:1212.2001}}.
\bibitem[{Aad et~al.(2013)}]{Aad:2012gla}
\bibinfo{author}{G.~Aad}, et~al. (\bibinfo{collaboration}{ATLAS}),
\newblock \bibinfo{title}{{Observation of Associated Near-Side and Away-Side
  Long-Range Correlations in $\sqrt{s_{NN}}$=5.02??TeV Proton-Lead Collisions
  with the ATLAS Detector}},
\newblock \bibinfo{journal}{Phys. Rev. Lett.} \bibinfo{volume}{110}
  (\bibinfo{year}{2013}) \bibinfo{pages}{182302}.
  \DOIprefix\doi{10.1103/PhysRevLett.110.182302}.
  \href{http://arxiv.org/abs/1212.5198}{{\tt arXiv:1212.5198}}.
\bibitem[{Aaij et~al.(2016)}]{Aaij:2015qcq}
\bibinfo{author}{R.~Aaij}, et~al. (\bibinfo{collaboration}{LHCb}),
\newblock \bibinfo{title}{{Measurements of long-range near-side angular
  correlations in $\sqrt{s_{\text{NN}}}=5$TeV proton-lead collisions in the
  forward region}},
\newblock \bibinfo{journal}{Phys. Lett.} \bibinfo{volume}{B762}
  (\bibinfo{year}{2016}) \bibinfo{pages}{473--483}.
  \DOIprefix\doi{10.1016/j.physletb.2016.09.064}.
  \href{http://arxiv.org/abs/1512.00439}{{\tt arXiv:1512.00439}}.
\bibitem[{Khachatryan et~al.(2017{\natexlab{a}})}]{Khachatryan:2016ibd}
\bibinfo{author}{V.~Khachatryan}, et~al. (\bibinfo{collaboration}{CMS}),
\newblock \bibinfo{title}{{Pseudorapidity dependence of long-range two-particle
  correlations in $p$Pb collisions at $\sqrt {s_{NN}}=$ 5.02 TeV}},
\newblock \bibinfo{journal}{Phys. Rev.} \bibinfo{volume}{C96}
  (\bibinfo{year}{2017}{\natexlab{a}}) \bibinfo{pages}{014915}.
  \DOIprefix\doi{10.1103/PhysRevC.96.014915}.
  \href{http://arxiv.org/abs/1604.05347}{{\tt arXiv:1604.05347}}.
\bibitem[{Khachatryan et~al.(2017{\natexlab{b}})}]{Khachatryan:2016txc}
\bibinfo{author}{V.~Khachatryan}, et~al. (\bibinfo{collaboration}{CMS}),
\newblock \bibinfo{title}{{Evidence for collectivity in pp collisions at the
  LHC}},
\newblock \bibinfo{journal}{Phys. Lett.} \bibinfo{volume}{B765}
  (\bibinfo{year}{2017}{\natexlab{b}}) \bibinfo{pages}{193--220}.
  \DOIprefix\doi{10.1016/j.physletb.2016.12.009}.
  \href{http://arxiv.org/abs/1606.06198}{{\tt arXiv:1606.06198}}.
\bibitem[{Aaboud et~al.(2017{\natexlab{a}})}]{Aaboud:2016yar}
\bibinfo{author}{M.~Aaboud}, et~al. (\bibinfo{collaboration}{ATLAS}),
\newblock \bibinfo{title}{{Measurements of long-range azimuthal anisotropies
  and associated Fourier coefficients for $pp$ collisions at $\sqrt{s}=5.02$
  and $13$ TeV and $p$+Pb collisions at $\sqrt{s_{\mathrm{NN}}}=5.02$ TeV with
  the ATLAS detector}},
\newblock \bibinfo{journal}{Phys. Rev.} \bibinfo{volume}{C96}
  (\bibinfo{year}{2017}{\natexlab{a}}) \bibinfo{pages}{024908}.
  \DOIprefix\doi{10.1103/PhysRevC.96.024908}.
  \href{http://arxiv.org/abs/1609.06213}{{\tt arXiv:1609.06213}}.
\bibitem[{Aaboud et~al.(2017{\natexlab{b}})}]{Aaboud:2017acw}
\bibinfo{author}{M.~Aaboud}, et~al. (\bibinfo{collaboration}{ATLAS}),
\newblock \bibinfo{title}{{Measurement of multi-particle azimuthal correlations
  in $pp$, $p+$Pb and low-multiplicity Pb$+$Pb collisions with the ATLAS
  detector}},
\newblock \bibinfo{journal}{Eur. Phys. J.} \bibinfo{volume}{C77}
  (\bibinfo{year}{2017}{\natexlab{b}}) \bibinfo{pages}{428}.
  \DOIprefix\doi{10.1140/epjc/s10052-017-4988-1}.
  \href{http://arxiv.org/abs/1705.04176}{{\tt arXiv:1705.04176}}.
\bibitem[{Aaboud et~al.(2018)}]{Aaboud:2017blb}
\bibinfo{author}{M.~Aaboud}, et~al. (\bibinfo{collaboration}{ATLAS}),
\newblock \bibinfo{title}{{Measurement of long-range multiparticle azimuthal
  correlations with the subevent cumulant method in $pp$ and $p + Pb$
  collisions with the ATLAS detector at the CERN Large Hadron Collider}},
\newblock \bibinfo{journal}{Phys. Rev.} \bibinfo{volume}{C97}
  (\bibinfo{year}{2018}) \bibinfo{pages}{024904}.
  \DOIprefix\doi{10.1103/PhysRevC.97.024904}.
  \href{http://arxiv.org/abs/1708.03559}{{\tt arXiv:1708.03559}}.
\bibitem[{Chatrchyan et~al.(2013)}]{Chatrchyan:2013nka}
\bibinfo{author}{S.~Chatrchyan}, et~al. (\bibinfo{collaboration}{CMS}),
\newblock \bibinfo{title}{{Multiplicity and transverse momentum dependence of
  two- and four-particle correlations in pPb and PbPb collisions}},
\newblock \bibinfo{journal}{Phys. Lett.} \bibinfo{volume}{B724}
  (\bibinfo{year}{2013}) \bibinfo{pages}{213--240}.
  \DOIprefix\doi{10.1016/j.physletb.2013.06.028}.
  \href{http://arxiv.org/abs/1305.0609}{{\tt arXiv:1305.0609}}.
\bibitem[{Abelev et~al.(2014)}]{Abelev:2014mda}
\bibinfo{author}{B.~B. Abelev}, et~al. (\bibinfo{collaboration}{ALICE}),
\newblock \bibinfo{title}{{Multiparticle azimuthal correlations in p -Pb and
  Pb-Pb collisions at the CERN Large Hadron Collider}},
\newblock \bibinfo{journal}{Phys. Rev.} \bibinfo{volume}{C90}
  (\bibinfo{year}{2014}) \bibinfo{pages}{054901}.
  \DOIprefix\doi{10.1103/PhysRevC.90.054901}.
  \href{http://arxiv.org/abs/1406.2474}{{\tt arXiv:1406.2474}}.
\bibitem[{Alver et~al.(2010)}]{Alver:2009id}
\bibinfo{author}{B.~Alver}, et~al. (\bibinfo{collaboration}{PHOBOS}),
\newblock \bibinfo{title}{{High transverse momentum triggered correlations over
  a large pseudorapidity acceptance in Au+Au collisions at s(NN)**1/2 = 200
  GeV}},
\newblock \bibinfo{journal}{Phys. Rev. Lett.} \bibinfo{volume}{104}
  (\bibinfo{year}{2010}) \bibinfo{pages}{062301}.
  \DOIprefix\doi{10.1103/PhysRevLett.104.062301}.
  \href{http://arxiv.org/abs/0903.2811}{{\tt arXiv:0903.2811}}.
\bibitem[{Abelev et~al.(2009)}]{Abelev:2009af}
\bibinfo{author}{B.~I. Abelev}, et~al. (\bibinfo{collaboration}{STAR}),
\newblock \bibinfo{title}{{Long range rapidity correlations and jet production
  in high energy nuclear collisions}},
\newblock \bibinfo{journal}{Phys. Rev.} \bibinfo{volume}{C80}
  (\bibinfo{year}{2009}) \bibinfo{pages}{064912}.
  \DOIprefix\doi{10.1103/PhysRevC.80.064912}.
  \href{http://arxiv.org/abs/0909.0191}{{\tt arXiv:0909.0191}}.
\bibitem[{Adare et~al.(2015)}]{Adare:2014keg}
\bibinfo{author}{A.~Adare}, et~al. (\bibinfo{collaboration}{PHENIX}),
\newblock \bibinfo{title}{{Measurement of long-range angular correlation and
  quadrupole anisotropy of pions and (anti)protons in central $d$$+$Au
  collisions at $\sqrt{s_{_{NN}}}$=200 GeV}},
\newblock \bibinfo{journal}{Phys. Rev. Lett.} \bibinfo{volume}{114}
  (\bibinfo{year}{2015}) \bibinfo{pages}{192301}.
  \DOIprefix\doi{10.1103/PhysRevLett.114.192301}.
  \href{http://arxiv.org/abs/1404.7461}{{\tt arXiv:1404.7461}}.
\bibitem[{Adamczyk et~al.(2015)}]{Adamczyk:2015xjc}
\bibinfo{author}{L.~Adamczyk}, et~al. (\bibinfo{collaboration}{STAR}),
\newblock \bibinfo{title}{{Long-range pseudorapidity dihadron correlations in
  $d$+Au collisions at $\sqrt{s_{\rm NN}}=200$ GeV}},
\newblock \bibinfo{journal}{Phys. Lett.} \bibinfo{volume}{B747}
  (\bibinfo{year}{2015}) \bibinfo{pages}{265--271}.
  \DOIprefix\doi{10.1016/j.physletb.2015.05.075}.
  \href{http://arxiv.org/abs/1502.07652}{{\tt arXiv:1502.07652}}.
\bibitem[{Adare et~al.(2015)}]{Adare:2015ctn}
\bibinfo{author}{A.~Adare}, et~al. (\bibinfo{collaboration}{PHENIX}),
\newblock \bibinfo{title}{{Measurements of elliptic and triangular flow in
  high-multiplicity $^{3}$He$+$Au collisions at $\sqrt{s_{_{NN}}}=200$ GeV}},
\newblock \bibinfo{journal}{Phys. Rev. Lett.} \bibinfo{volume}{115}
  (\bibinfo{year}{2015}) \bibinfo{pages}{142301}.
  \DOIprefix\doi{10.1103/PhysRevLett.115.142301}.
  \href{http://arxiv.org/abs/1507.06273}{{\tt arXiv:1507.06273}}.
\bibitem[{Blok and Wiedemann(2019)}]{Blok:2018xes}
\bibinfo{author}{B.~Blok}, \bibinfo{author}{U.~A. Wiedemann},
\newblock \bibinfo{title}{{Collectivity in pp from resummed interference
  effects?}},
\newblock \bibinfo{journal}{Phys. Lett.} \bibinfo{volume}{B795}
  (\bibinfo{year}{2019}) \bibinfo{pages}{259--265}.
  \DOIprefix\doi{10.1016/j.physletb.2019.05.038}.
  \href{http://arxiv.org/abs/1812.04113}{{\tt arXiv:1812.04113}}.
\bibitem[{Kurkela et~al.(2019)Kurkela, Wiedemann, and Wu}]{Kurkela:2019kip}
\bibinfo{author}{A.~Kurkela}, \bibinfo{author}{U.~A. Wiedemann},
  \bibinfo{author}{B.~Wu},
\newblock \bibinfo{title}{{Flow in AA and pA as an interplay of fluid-like and
  non-fluid like excitations}}  (\bibinfo{year}{2019}).
  \href{http://arxiv.org/abs/1905.05139}{{\tt arXiv:1905.05139}}.
\bibitem[{Nie et~al.(2019)Nie, Yi, Jia, and Ma}]{Nie:2019swk}
\bibinfo{author}{M.~Nie}, \bibinfo{author}{L.~Yi}, \bibinfo{author}{J.~Jia},
  \bibinfo{author}{G.~Ma},
\newblock \bibinfo{title}{{Influence of initial-state momentum anisotropy on
  the final-state collectivity in small collision systems}}
  (\bibinfo{year}{2019}). \href{http://arxiv.org/abs/1906.01422}{{\tt
  arXiv:1906.01422}}.
\bibitem[{Kovchegov and Levin(2012)}]{Kovchegov:2012mbw}
\bibinfo{author}{Y.~V. Kovchegov}, \bibinfo{author}{E.~Levin},
  \bibinfo{title}{{Quantum chromodynamics at high energy}},
  volume~\bibinfo{volume}{33}, \bibinfo{publisher}{Cambridge University Press},
  \bibinfo{year}{2012}. \URLprefix
  \url{http://www.cambridge.org/de/knowledge/isbn/item6803159}.
\bibitem[{Iancu et~al.(2002)Iancu, Leonidov, and McLerran}]{Iancu:2002xk}
\bibinfo{author}{E.~Iancu}, \bibinfo{author}{A.~Leonidov},
  \bibinfo{author}{L.~McLerran},
\newblock \bibinfo{title}{{The Color glass condensate: An Introduction}},
\newblock in: \bibinfo{booktitle}{{QCD perspectives on hot and dense matter.
  Proceedings, NATO Advanced Study Institute, Summer School, Cargese, France,
  August 6-18, 2001}}, \bibinfo{year}{2002}, pp. \bibinfo{pages}{73--145}.
  \href{http://arxiv.org/abs/hep-ph/0202270}{{\tt arXiv:hep-ph/0202270}}.
\bibitem[{McLerran(2008)}]{McLerran:2008uj}
\bibinfo{author}{L.~McLerran},
\newblock \bibinfo{title}{{The Color Glass Condensate and Glasma}}
  (\bibinfo{year}{2008}). \href{http://arxiv.org/abs/0804.1736}{{\tt
  arXiv:0804.1736}}.
\bibitem[{Gelis et~al.(2010)Gelis, Iancu, Jalilian-Marian, and
  Venugopalan}]{Gelis:2010nm}
\bibinfo{author}{F.~Gelis}, \bibinfo{author}{E.~Iancu},
  \bibinfo{author}{J.~Jalilian-Marian}, \bibinfo{author}{R.~Venugopalan},
\newblock \bibinfo{title}{{The Color Glass Condensate}},
\newblock \bibinfo{journal}{Ann. Rev. Nucl. Part. Sci.} \bibinfo{volume}{60}
  (\bibinfo{year}{2010}) \bibinfo{pages}{463--489}.
  \DOIprefix\doi{10.1146/annurev.nucl.010909.083629}.
  \href{http://arxiv.org/abs/1002.0333}{{\tt arXiv:1002.0333}}.
\bibitem[{Dumitru et~al.(2008)Dumitru, Gelis, McLerran, and
  Venugopalan}]{Dumitru:2008wn}
\bibinfo{author}{A.~Dumitru}, \bibinfo{author}{F.~Gelis},
  \bibinfo{author}{L.~McLerran}, \bibinfo{author}{R.~Venugopalan},
\newblock \bibinfo{title}{{Glasma flux tubes and the near side ridge phenomenon
  at RHIC}},
\newblock \bibinfo{journal}{Nucl. Phys.} \bibinfo{volume}{A810}
  (\bibinfo{year}{2008}) \bibinfo{pages}{91--108}.
  \DOIprefix\doi{10.1016/j.nuclphysa.2008.06.012}.
  \href{http://arxiv.org/abs/0804.3858}{{\tt arXiv:0804.3858}}.
\bibitem[{Dumitru et~al.(2011)Dumitru, Dusling, Gelis, Jalilian-Marian, Lappi,
  and Venugopalan}]{Dumitru:2010iy}
\bibinfo{author}{A.~Dumitru}, \bibinfo{author}{K.~Dusling},
  \bibinfo{author}{F.~Gelis}, \bibinfo{author}{J.~Jalilian-Marian},
  \bibinfo{author}{T.~Lappi}, \bibinfo{author}{R.~Venugopalan},
\newblock \bibinfo{title}{{The Ridge in proton-proton collisions at the LHC}},
\newblock \bibinfo{journal}{Phys. Lett.} \bibinfo{volume}{B697}
  (\bibinfo{year}{2011}) \bibinfo{pages}{21--25}.
  \DOIprefix\doi{10.1016/j.physletb.2011.01.024}.
  \href{http://arxiv.org/abs/1009.5295}{{\tt arXiv:1009.5295}}.
\bibitem[{Kovchegov and Wertepny(2013)}]{Kovchegov:2012nd}
\bibinfo{author}{Y.~V. Kovchegov}, \bibinfo{author}{D.~E. Wertepny},
\newblock \bibinfo{title}{{Long-Range Rapidity Correlations in Heavy-Light Ion
  Collisions}},
\newblock \bibinfo{journal}{Nucl. Phys.} \bibinfo{volume}{A906}
  (\bibinfo{year}{2013}) \bibinfo{pages}{50--83}.
  \DOIprefix\doi{10.1016/j.nuclphysa.2013.03.006}.
  \href{http://arxiv.org/abs/1212.1195}{{\tt arXiv:1212.1195}}.
\bibitem[{Kovchegov and Wertepny(2014)}]{Kovchegov:2013ewa}
\bibinfo{author}{Y.~V. Kovchegov}, \bibinfo{author}{D.~E. Wertepny},
\newblock \bibinfo{title}{{Two-Gluon Correlations in Heavy-Light Ion
  Collisions: Energy and Geometry Dependence, IR Divergences, and
  $k_T$-Factorization}},
\newblock \bibinfo{journal}{Nucl. Phys.} \bibinfo{volume}{A925}
  (\bibinfo{year}{2014}) \bibinfo{pages}{254--295}.
  \DOIprefix\doi{10.1016/j.nuclphysa.2014.02.021}.
  \href{http://arxiv.org/abs/1310.6701}{{\tt arXiv:1310.6701}}.
\bibitem[{Altinoluk et~al.(2015)Altinoluk, Armesto, Beuf, Kovner, and
  Lublinsky}]{Altinoluk:2015uaa}
\bibinfo{author}{T.~Altinoluk}, \bibinfo{author}{N.~Armesto},
  \bibinfo{author}{G.~Beuf}, \bibinfo{author}{A.~Kovner},
  \bibinfo{author}{M.~Lublinsky},
\newblock \bibinfo{title}{{Bose enhancement and the ridge}},
\newblock \bibinfo{journal}{Phys. Lett.} \bibinfo{volume}{B751}
  (\bibinfo{year}{2015}) \bibinfo{pages}{448--452}.
  \DOIprefix\doi{10.1016/j.physletb.2015.10.072}.
  \href{http://arxiv.org/abs/1503.07126}{{\tt arXiv:1503.07126}}.
\bibitem[{Altinoluk et~al.(2016)Altinoluk, Armesto, Beuf, Kovner, and
  Lublinsky}]{Altinoluk:2015eka}
\bibinfo{author}{T.~Altinoluk}, \bibinfo{author}{N.~Armesto},
  \bibinfo{author}{G.~Beuf}, \bibinfo{author}{A.~Kovner},
  \bibinfo{author}{M.~Lublinsky},
\newblock \bibinfo{title}{{Hanbury?Brown?Twiss measurements at large rapidity
  separations, or can we measure the proton radius in p-A collisions?}},
\newblock \bibinfo{journal}{Phys. Lett.} \bibinfo{volume}{B752}
  (\bibinfo{year}{2016}) \bibinfo{pages}{113--121}.
  \DOIprefix\doi{10.1016/j.physletb.2015.11.033}.
  \href{http://arxiv.org/abs/1509.03223}{{\tt arXiv:1509.03223}}.
\bibitem[{Dusling and Venugopalan(2012)}]{Dusling:2012iga}
\bibinfo{author}{K.~Dusling}, \bibinfo{author}{R.~Venugopalan},
\newblock \bibinfo{title}{{Azimuthal collimation of long range rapidity
  correlations by strong color fields in high multiplicity hadron-hadron
  collisions}},
\newblock \bibinfo{journal}{Phys. Rev. Lett.} \bibinfo{volume}{108}
  (\bibinfo{year}{2012}) \bibinfo{pages}{262001}.
  \DOIprefix\doi{10.1103/PhysRevLett.108.262001}.
  \href{http://arxiv.org/abs/1201.2658}{{\tt arXiv:1201.2658}}.
\bibitem[{Dusling and Venugopalan(2013{\natexlab{a}})}]{Dusling:2012cg}
\bibinfo{author}{K.~Dusling}, \bibinfo{author}{R.~Venugopalan},
\newblock \bibinfo{title}{{Evidence for BFKL and saturation dynamics from
  dihadron spectra at the LHC}},
\newblock \bibinfo{journal}{Phys. Rev.} \bibinfo{volume}{D87}
  (\bibinfo{year}{2013}{\natexlab{a}}) \bibinfo{pages}{051502}.
  \DOIprefix\doi{10.1103/PhysRevD.87.051502}.
  \href{http://arxiv.org/abs/1210.3890}{{\tt arXiv:1210.3890}}.
\bibitem[{Dusling and Venugopalan(2013{\natexlab{b}})}]{Dusling:2012wy}
\bibinfo{author}{K.~Dusling}, \bibinfo{author}{R.~Venugopalan},
\newblock \bibinfo{title}{{Explanation of systematics of CMS p+Pb high
  multiplicity di-hadron data at $\sqrt{s}_{\rm NN} = 5.02$ TeV}},
\newblock \bibinfo{journal}{Phys. Rev.} \bibinfo{volume}{D87}
  (\bibinfo{year}{2013}{\natexlab{b}}) \bibinfo{pages}{054014}.
  \DOIprefix\doi{10.1103/PhysRevD.87.054014}.
  \href{http://arxiv.org/abs/1211.3701}{{\tt arXiv:1211.3701}}.
\bibitem[{Dusling and Venugopalan(2013{\natexlab{c}})}]{Dusling:2013qoz}
\bibinfo{author}{K.~Dusling}, \bibinfo{author}{R.~Venugopalan},
\newblock \bibinfo{title}{{Comparison of the color glass condensate to dihadron
  correlations in proton-proton and proton-nucleus collisions}},
\newblock \bibinfo{journal}{Phys. Rev.} \bibinfo{volume}{D87}
  (\bibinfo{year}{2013}{\natexlab{c}}) \bibinfo{pages}{094034}.
  \DOIprefix\doi{10.1103/PhysRevD.87.094034}.
  \href{http://arxiv.org/abs/1302.7018}{{\tt arXiv:1302.7018}}.
\bibitem[{Ozonder(2015)}]{Ozonder:2014sra}
\bibinfo{author}{S.~Ozonder},
\newblock \bibinfo{title}{{Triple-gluon and quadruple-gluon azimuthal
  correlations from glasma and higher-dimensional ridges}},
\newblock \bibinfo{journal}{Phys. Rev.} \bibinfo{volume}{D91}
  (\bibinfo{year}{2015}) \bibinfo{pages}{034005}.
  \DOIprefix\doi{10.1103/PhysRevD.91.034005}.
  \href{http://arxiv.org/abs/1409.6347}{{\tt arXiv:1409.6347}}.
\bibitem[{Ozonder(2018)}]{Ozonder:2017wmh}
\bibinfo{author}{S.~Ozonder},
\newblock \bibinfo{title}{{Predictions on three-particle azimuthal correlations
  in proton-proton collisions}},
\newblock \bibinfo{journal}{Turk. J. Phys.} \bibinfo{volume}{42}
  (\bibinfo{year}{2018}) \bibinfo{pages}{78--83}.
  \DOIprefix\doi{10.3906/fiz-1710-6}.
  \href{http://arxiv.org/abs/1712.05571}{{\tt arXiv:1712.05571}}.
\bibitem[{Altinoluk et~al.(2017)Altinoluk, Armesto, Beuf, Kovner, and
  Lublinsky}]{Altinoluk:2016vax}
\bibinfo{author}{T.~Altinoluk}, \bibinfo{author}{N.~Armesto},
  \bibinfo{author}{G.~Beuf}, \bibinfo{author}{A.~Kovner},
  \bibinfo{author}{M.~Lublinsky},
\newblock \bibinfo{title}{{Quark correlations in the Color Glass Condensate:
  Pauli blocking and the ridge}},
\newblock \bibinfo{journal}{Phys. Rev.} \bibinfo{volume}{D95}
  (\bibinfo{year}{2017}) \bibinfo{pages}{034025}.
  \DOIprefix\doi{10.1103/PhysRevD.95.034025}.
  \href{http://arxiv.org/abs/1610.03020}{{\tt arXiv:1610.03020}}.
\bibitem[{Martinez et~al.(2018)Martinez, Sievert, and
  Wertepny}]{Martinez:2018ygo}
\bibinfo{author}{M.~Martinez}, \bibinfo{author}{M.~D. Sievert},
  \bibinfo{author}{D.~E. Wertepny},
\newblock \bibinfo{title}{{Toward Initial Conditions of Conserved Charges Part
  I: Spatial Correlations of Quarks and Antiquarks}},
\newblock \bibinfo{journal}{JHEP} \bibinfo{volume}{07} (\bibinfo{year}{2018})
  \bibinfo{pages}{003}. \DOIprefix\doi{10.1007/JHEP07(2018)003}.
  \href{http://arxiv.org/abs/1801.08986}{{\tt arXiv:1801.08986}}.
\bibitem[{Lappi et~al.(2016)Lappi, Schenke, Schlichting, and
  Venugopalan}]{Lappi:2015vta}
\bibinfo{author}{T.~Lappi}, \bibinfo{author}{B.~Schenke},
  \bibinfo{author}{S.~Schlichting}, \bibinfo{author}{R.~Venugopalan},
\newblock \bibinfo{title}{{Tracing the origin of azimuthal gluon correlations
  in the color glass condensate}},
\newblock \bibinfo{journal}{JHEP} \bibinfo{volume}{01} (\bibinfo{year}{2016})
  \bibinfo{pages}{061}. \DOIprefix\doi{10.1007/JHEP01(2016)061}.
  \href{http://arxiv.org/abs/1509.03499}{{\tt arXiv:1509.03499}}.
\bibitem[{Altinoluk et~al.(2018{\natexlab{a}})Altinoluk, Armesto, and
  Wertepny}]{Altinoluk:2018hcu}
\bibinfo{author}{T.~Altinoluk}, \bibinfo{author}{N.~Armesto},
  \bibinfo{author}{D.~E. Wertepny},
\newblock \bibinfo{title}{{Correlations and the ridge in the Color Glass
  Condensate beyond the glasma graph approximation}},
\newblock \bibinfo{journal}{JHEP} \bibinfo{volume}{05}
  (\bibinfo{year}{2018}{\natexlab{a}}) \bibinfo{pages}{207}.
  \DOIprefix\doi{10.1007/JHEP05(2018)207}.
  \href{http://arxiv.org/abs/1804.02910}{{\tt arXiv:1804.02910}}.
\bibitem[{Altinoluk et~al.(2018{\natexlab{b}})Altinoluk, Armesto, Kovner, and
  Lublinsky}]{Altinoluk:2018ogz}
\bibinfo{author}{T.~Altinoluk}, \bibinfo{author}{N.~Armesto},
  \bibinfo{author}{A.~Kovner}, \bibinfo{author}{M.~Lublinsky},
\newblock \bibinfo{title}{{Double and triple inclusive gluon production at mid
  rapidity: quantum interference in p-A scattering}},
\newblock \bibinfo{journal}{Eur. Phys. J.} \bibinfo{volume}{C78}
  (\bibinfo{year}{2018}{\natexlab{b}}) \bibinfo{pages}{702}.
  \DOIprefix\doi{10.1140/epjc/s10052-018-6186-1}.
  \href{http://arxiv.org/abs/1805.07739}{{\tt arXiv:1805.07739}}.
\bibitem[{Martinez et~al.(2019)Martinez, Sievert, and
  Wertepny}]{Martinez:2018tuf}
\bibinfo{author}{M.~Martinez}, \bibinfo{author}{M.~D. Sievert},
  \bibinfo{author}{D.~E. Wertepny},
\newblock \bibinfo{title}{{Multiparticle Production at Mid-Rapidity in the
  Color-Glass Condensate}},
\newblock \bibinfo{journal}{JHEP} \bibinfo{volume}{02} (\bibinfo{year}{2019})
  \bibinfo{pages}{024}. \DOIprefix\doi{10.1007/JHEP02(2019)024}.
  \href{http://arxiv.org/abs/1808.04896}{{\tt arXiv:1808.04896}}.
\bibitem[{Dusling et~al.(2018{\natexlab{a}})Dusling, Mace, and
  Venugopalan}]{Dusling:2017dqg}
\bibinfo{author}{K.~Dusling}, \bibinfo{author}{M.~Mace},
  \bibinfo{author}{R.~Venugopalan},
\newblock \bibinfo{title}{{Multiparticle collectivity from initial state
  correlations in high energy proton-nucleus collisions}},
\newblock \bibinfo{journal}{Phys. Rev. Lett.} \bibinfo{volume}{120}
  (\bibinfo{year}{2018}{\natexlab{a}}) \bibinfo{pages}{042002}.
  \DOIprefix\doi{10.1103/PhysRevLett.120.042002}.
  \href{http://arxiv.org/abs/1705.00745}{{\tt arXiv:1705.00745}}.
\bibitem[{Dusling et~al.(2018{\natexlab{b}})Dusling, Mace, and
  Venugopalan}]{Dusling:2017aot}
\bibinfo{author}{K.~Dusling}, \bibinfo{author}{M.~Mace},
  \bibinfo{author}{R.~Venugopalan},
\newblock \bibinfo{title}{{Parton model description of multiparticle azimuthal
  correlations in $pA$ collisions}},
\newblock \bibinfo{journal}{Phys. Rev.} \bibinfo{volume}{D97}
  (\bibinfo{year}{2018}{\natexlab{b}}) \bibinfo{pages}{016014}.
  \DOIprefix\doi{10.1103/PhysRevD.97.016014}.
  \href{http://arxiv.org/abs/1706.06260}{{\tt arXiv:1706.06260}}.
\bibitem[{Levin and Rezaeian(2011)}]{Levin:2011fb}
\bibinfo{author}{E.~Levin}, \bibinfo{author}{A.~H. Rezaeian},
\newblock \bibinfo{title}{{The Ridge from the BFKL evolution and beyond}},
\newblock \bibinfo{journal}{Phys. Rev.} \bibinfo{volume}{D84}
  (\bibinfo{year}{2011}) \bibinfo{pages}{034031}.
  \DOIprefix\doi{10.1103/PhysRevD.84.034031}.
  \href{http://arxiv.org/abs/1105.3275}{{\tt arXiv:1105.3275}}.
\bibitem[{McLerran and Skokov(2017)}]{McLerran:2016snu}
\bibinfo{author}{L.~McLerran}, \bibinfo{author}{V.~Skokov},
\newblock \bibinfo{title}{{Odd Azimuthal Anisotropy of the Glasma for pA
  Scattering}},
\newblock \bibinfo{journal}{Nucl. Phys.} \bibinfo{volume}{A959}
  (\bibinfo{year}{2017}) \bibinfo{pages}{83--101}.
  \DOIprefix\doi{10.1016/j.nuclphysa.2016.12.011}.
  \href{http://arxiv.org/abs/1611.09870}{{\tt arXiv:1611.09870}}.
\bibitem[{Kovner et~al.(2017)Kovner, Lublinsky, and Skokov}]{Kovner:2016jfp}
\bibinfo{author}{A.~Kovner}, \bibinfo{author}{M.~Lublinsky},
  \bibinfo{author}{V.~Skokov},
\newblock \bibinfo{title}{{Exploring correlations in the CGC wave function: odd
  azimuthal anisotropy}},
\newblock \bibinfo{journal}{Phys. Rev.} \bibinfo{volume}{D96}
  (\bibinfo{year}{2017}) \bibinfo{pages}{016010}.
  \DOIprefix\doi{10.1103/PhysRevD.96.016010}.
  \href{http://arxiv.org/abs/1612.07790}{{\tt arXiv:1612.07790}}.
\bibitem[{Kovchegov and Skokov(2018)}]{Kovchegov:2018jun}
\bibinfo{author}{Y.~V. Kovchegov}, \bibinfo{author}{V.~V. Skokov},
\newblock \bibinfo{title}{{How classical gluon fields generate odd azimuthal
  harmonics for the two-gluon correlation function in high-energy collisions}},
\newblock \bibinfo{journal}{Phys. Rev.} \bibinfo{volume}{D97}
  (\bibinfo{year}{2018}) \bibinfo{pages}{094021}.
  \DOIprefix\doi{10.1103/PhysRevD.97.094021}.
  \href{http://arxiv.org/abs/1802.08166}{{\tt arXiv:1802.08166}}.
\bibitem[{Mace et~al.(2018)Mace, Skokov, Tribedy, and
  Venugopalan}]{Mace:2018vwq}
\bibinfo{author}{M.~Mace}, \bibinfo{author}{V.~V. Skokov},
  \bibinfo{author}{P.~Tribedy}, \bibinfo{author}{R.~Venugopalan},
\newblock \bibinfo{title}{{Hierarchy of azimuthal anisotropy harmonics in
  collisions of small systems from the Color Glass Condensate}},
\newblock \bibinfo{journal}{Phys. Rev. Lett.} \bibinfo{volume}{121}
  (\bibinfo{year}{2018}) \bibinfo{pages}{052301}.
  \DOIprefix\doi{10.1103/PhysRevLett.121.052301}.
  \href{http://arxiv.org/abs/1805.09342}{{\tt arXiv:1805.09342}}.
\bibitem[{Mace et~al.(2019{\natexlab{a}})Mace, Skokov, Tribedy, and
  Venugopalan}]{Mace:2018yvl}
\bibinfo{author}{M.~Mace}, \bibinfo{author}{V.~V. Skokov},
  \bibinfo{author}{P.~Tribedy}, \bibinfo{author}{R.~Venugopalan},
\newblock \bibinfo{title}{{Systematics of azimuthal anisotropy harmonics in
  proton?nucleus collisions at the LHC from the Color Glass Condensate}},
\newblock \bibinfo{journal}{Phys. Lett.} \bibinfo{volume}{B788}
  (\bibinfo{year}{2019}{\natexlab{a}}) \bibinfo{pages}{161--165}.
  \DOIprefix\doi{10.1016/j.physletb.2018.09.064}.
  \href{http://arxiv.org/abs/1807.00825}{{\tt arXiv:1807.00825}}.
\bibitem[{Mace et~al.(2019{\natexlab{b}})Mace, Skokov, Tribedy, and
  Venugopalan}]{Mace:2019rtt}
\bibinfo{author}{M.~Mace}, \bibinfo{author}{V.~V. Skokov},
  \bibinfo{author}{P.~Tribedy}, \bibinfo{author}{R.~Venugopalan},
\newblock \bibinfo{title}{{Initial state description of azimuthally collimated
  long range correlations in ultrarelativistic light-heavy ion collisions}}
  (\bibinfo{year}{2019}{\natexlab{b}}).
  \href{http://arxiv.org/abs/1901.10506}{{\tt arXiv:1901.10506}}.
\bibitem[{Dumitru and Skokov(2015)}]{Dumitru:2014vka}
\bibinfo{author}{A.~Dumitru}, \bibinfo{author}{V.~Skokov},
\newblock \bibinfo{title}{{Anisotropy of the semiclassical gluon field of a
  large nucleus at high energy}},
\newblock \bibinfo{journal}{Phys. Rev.} \bibinfo{volume}{D91}
  (\bibinfo{year}{2015}) \bibinfo{pages}{074006}.
  \DOIprefix\doi{10.1103/PhysRevD.91.074006}.
  \href{http://arxiv.org/abs/1411.6630}{{\tt arXiv:1411.6630}}.
\bibitem[{Kovner et~al.(2017)Kovner, Lublinsky, and Skokov}]{Kovner:2017gab}
\bibinfo{author}{A.~Kovner}, \bibinfo{author}{M.~Lublinsky},
  \bibinfo{author}{V.~Skokov},
\newblock \bibinfo{title}{{Initial state qqg correlations as a background for
  the Chiral Magnetic Effect in collision of small systems}},
\newblock \bibinfo{journal}{Phys. Rev.} \bibinfo{volume}{D96}
  (\bibinfo{year}{2017}) \bibinfo{pages}{096003}.
  \DOIprefix\doi{10.1103/PhysRevD.96.096003}.
  \href{http://arxiv.org/abs/1706.02330}{{\tt arXiv:1706.02330}}.
\bibitem[{Davy et~al.(2019)Davy, Marquet, Shi, Xiao, and Zhang}]{Davy:2018hsl}
\bibinfo{author}{M.~K. Davy}, \bibinfo{author}{C.~Marquet},
  \bibinfo{author}{Y.~Shi}, \bibinfo{author}{B.-W. Xiao},
  \bibinfo{author}{C.~Zhang},
\newblock \bibinfo{title}{{Two particle azimuthal harmonics in pA collisions}},
\newblock \bibinfo{journal}{Nucl. Phys.} \bibinfo{volume}{A983}
  (\bibinfo{year}{2019}) \bibinfo{pages}{293--309}.
  \DOIprefix\doi{10.1016/j.nuclphysa.2018.11.001}.
  \href{http://arxiv.org/abs/1808.09851}{{\tt arXiv:1808.09851}}.
\bibitem[{Kovner and Lublinsky(2013)}]{Kovner:2012jm}
\bibinfo{author}{A.~Kovner}, \bibinfo{author}{M.~Lublinsky},
\newblock \bibinfo{title}{{Angular and long range rapidity correlations in
  particle production at high energy}},
\newblock \bibinfo{journal}{Int. J. Mod. Phys.} \bibinfo{volume}{E22}
  (\bibinfo{year}{2013}) \bibinfo{pages}{1330001}.
  \DOIprefix\doi{10.1142/S0218301313300014}.
  \href{http://arxiv.org/abs/1211.1928}{{\tt arXiv:1211.1928}}.
\bibitem[{Dumitru et~al.(2015)Dumitru, McLerran, and Skokov}]{Dumitru:2014yza}
\bibinfo{author}{A.~Dumitru}, \bibinfo{author}{L.~McLerran},
  \bibinfo{author}{V.~Skokov},
\newblock \bibinfo{title}{{Azimuthal asymmetries and the emergence of
  ?collectivity? from multi-particle correlations in high-energy pA
  collisions}},
\newblock \bibinfo{journal}{Phys. Lett.} \bibinfo{volume}{B743}
  (\bibinfo{year}{2015}) \bibinfo{pages}{134--137}.
  \DOIprefix\doi{10.1016/j.physletb.2015.02.046}.
  \href{http://arxiv.org/abs/1410.4844}{{\tt arXiv:1410.4844}}.
\bibitem[{McLerran and Venugopalan(1994{\natexlab{a}})}]{McLerran:1993ni}
\bibinfo{author}{L.~D. McLerran}, \bibinfo{author}{R.~Venugopalan},
\newblock \bibinfo{title}{{Computing quark and gluon distribution functions for
  very large nuclei}},
\newblock \bibinfo{journal}{Phys. Rev.} \bibinfo{volume}{D49}
  (\bibinfo{year}{1994}{\natexlab{a}}) \bibinfo{pages}{2233--2241}.
  \DOIprefix\doi{10.1103/PhysRevD.49.2233}.
  \href{http://arxiv.org/abs/hep-ph/9309289}{{\tt arXiv:hep-ph/9309289}}.
\bibitem[{McLerran and Venugopalan(1994{\natexlab{b}})}]{McLerran:1994vd}
\bibinfo{author}{L.~D. McLerran}, \bibinfo{author}{R.~Venugopalan},
\newblock \bibinfo{title}{{Green's functions in the color field of a large
  nucleus}},
\newblock \bibinfo{journal}{Phys. Rev.} \bibinfo{volume}{D50}
  (\bibinfo{year}{1994}{\natexlab{b}}) \bibinfo{pages}{2225--2233}.
  \DOIprefix\doi{10.1103/PhysRevD.50.2225}.
  \href{http://arxiv.org/abs/hep-ph/9402335}{{\tt arXiv:hep-ph/9402335}}.
\bibitem[{Kovner and Wiedemann(2003)}]{Kovner:2003zj}
\bibinfo{author}{A.~Kovner}, \bibinfo{author}{U.~A. Wiedemann},
\newblock \bibinfo{title}{{Gluon radiation and parton energy loss}}
  (\bibinfo{year}{2003}) \bibinfo{pages}{192--248}.
  \DOIprefix\doi{10.1142/9789812795533_0004}.
  \href{http://arxiv.org/abs/hep-ph/0304151}{{\tt arXiv:hep-ph/0304151}}.
\bibitem[{Casalderrey-Solana and Salgado(2007)}]{CasalderreySolana:2007pr}
\bibinfo{author}{J.~Casalderrey-Solana}, \bibinfo{author}{C.~A. Salgado},
\newblock \bibinfo{title}{{Introductory lectures on jet quenching in heavy ion
  collisions}},
\newblock \bibinfo{journal}{Acta Phys. Polon.} \bibinfo{volume}{B38}
  (\bibinfo{year}{2007}) \bibinfo{pages}{3731--3794}.
  \href{http://arxiv.org/abs/0712.3443}{{\tt arXiv:0712.3443}}.
\bibitem[{Balitsky and Tarasov(2015)}]{Balitsky:2015qba}
\bibinfo{author}{I.~Balitsky}, \bibinfo{author}{A.~Tarasov},
\newblock \bibinfo{title}{{Rapidity evolution of gluon TMD from low to moderate
  x}},
\newblock \bibinfo{journal}{JHEP} \bibinfo{volume}{10} (\bibinfo{year}{2015})
  \bibinfo{pages}{017}. \DOIprefix\doi{10.1007/JHEP10(2015)017}.
  \href{http://arxiv.org/abs/1505.02151}{{\tt arXiv:1505.02151}}.
\bibitem[{Balitsky and Tarasov(2016)}]{Balitsky:2016dgz}
\bibinfo{author}{I.~Balitsky}, \bibinfo{author}{A.~Tarasov},
\newblock \bibinfo{title}{{Gluon TMD in particle production from low to
  moderate x}},
\newblock \bibinfo{journal}{JHEP} \bibinfo{volume}{06} (\bibinfo{year}{2016})
  \bibinfo{pages}{164}. \DOIprefix\doi{10.1007/JHEP06(2016)164}.
  \href{http://arxiv.org/abs/1603.06548}{{\tt arXiv:1603.06548}}.
\bibitem[{Kovchegov et~al.(2016)Kovchegov, Pitonyak, and
  Sievert}]{Kovchegov:2015pbl}
\bibinfo{author}{Y.~V. Kovchegov}, \bibinfo{author}{D.~Pitonyak},
  \bibinfo{author}{M.~D. Sievert},
\newblock \bibinfo{title}{{Helicity Evolution at Small-x}},
\newblock \bibinfo{journal}{JHEP} \bibinfo{volume}{01} (\bibinfo{year}{2016})
  \bibinfo{pages}{072}. \DOIprefix\doi{10.1007/JHEP01(2016)072,
  10.1007/JHEP10(2016)148}. \href{http://arxiv.org/abs/1511.06737}{{\tt
  arXiv:1511.06737}}, \bibinfo{note}{[Erratum: JHEP10,148(2016)]}.
\bibitem[{Kovchegov et~al.(2017{\natexlab{a}})Kovchegov, Pitonyak, and
  Sievert}]{Kovchegov:2016zex}
\bibinfo{author}{Y.~V. Kovchegov}, \bibinfo{author}{D.~Pitonyak},
  \bibinfo{author}{M.~D. Sievert},
\newblock \bibinfo{title}{{Helicity Evolution at Small $x$: Flavor Singlet and
  Non-Singlet Observables}},
\newblock \bibinfo{journal}{Phys. Rev.} \bibinfo{volume}{D95}
  (\bibinfo{year}{2017}{\natexlab{a}}) \bibinfo{pages}{014033}.
  \DOIprefix\doi{10.1103/PhysRevD.95.014033}.
  \href{http://arxiv.org/abs/1610.06197}{{\tt arXiv:1610.06197}}.
\bibitem[{Kovchegov et~al.(2017{\natexlab{b}})Kovchegov, Pitonyak, and
  Sievert}]{Kovchegov:2017jxc}
\bibinfo{author}{Y.~V. Kovchegov}, \bibinfo{author}{D.~Pitonyak},
  \bibinfo{author}{M.~D. Sievert},
\newblock \bibinfo{title}{{Small-$x$ Asymptotics of the Quark Helicity
  Distribution: Analytic Results}},
\newblock \bibinfo{journal}{Phys. Lett.} \bibinfo{volume}{B772}
  (\bibinfo{year}{2017}{\natexlab{b}}) \bibinfo{pages}{136--140}.
  \DOIprefix\doi{10.1016/j.physletb.2017.06.032}.
  \href{http://arxiv.org/abs/1703.05809}{{\tt arXiv:1703.05809}}.
\bibitem[{Kovchegov et~al.(2017{\natexlab{c}})Kovchegov, Pitonyak, and
  Sievert}]{Kovchegov:2017lsr}
\bibinfo{author}{Y.~V. Kovchegov}, \bibinfo{author}{D.~Pitonyak},
  \bibinfo{author}{M.~D. Sievert},
\newblock \bibinfo{title}{{Small-$x$ Asymptotics of the Gluon Helicity
  Distribution}},
\newblock \bibinfo{journal}{JHEP} \bibinfo{volume}{10}
  (\bibinfo{year}{2017}{\natexlab{c}}) \bibinfo{pages}{198}.
  \DOIprefix\doi{10.1007/JHEP10(2017)198}.
  \href{http://arxiv.org/abs/1706.04236}{{\tt arXiv:1706.04236}}.
\bibitem[{Chirilli(2019)}]{Chirilli:2018kkw}
\bibinfo{author}{G.~A. Chirilli},
\newblock \bibinfo{title}{{Sub-eikonal corrections to scattering amplitudes at
  high energy}},
\newblock \bibinfo{journal}{JHEP} \bibinfo{volume}{01} (\bibinfo{year}{2019})
  \bibinfo{pages}{118}. \DOIprefix\doi{10.1007/JHEP01(2019)118}.
  \href{http://arxiv.org/abs/1807.11435}{{\tt arXiv:1807.11435}}.
\bibitem[{Laenen et~al.(2008)Laenen, Magnea, and Stavenga}]{Laenen:2008ux}
\bibinfo{author}{E.~Laenen}, \bibinfo{author}{L.~Magnea},
  \bibinfo{author}{G.~Stavenga},
\newblock \bibinfo{title}{{On next-to-eikonal corrections to threshold
  resummation for the Drell-Yan and DIS cross sections}},
\newblock \bibinfo{journal}{Phys. Lett.} \bibinfo{volume}{B669}
  (\bibinfo{year}{2008}) \bibinfo{pages}{173--179}.
  \DOIprefix\doi{10.1016/j.physletb.2008.09.037}.
  \href{http://arxiv.org/abs/0807.4412}{{\tt arXiv:0807.4412}}.
\bibitem[{Laenen et~al.(2009)Laenen, Stavenga, and White}]{Laenen:2008gt}
\bibinfo{author}{E.~Laenen}, \bibinfo{author}{G.~Stavenga},
  \bibinfo{author}{C.~D. White},
\newblock \bibinfo{title}{{Path integral approach to eikonal and
  next-to-eikonal exponentiation}},
\newblock \bibinfo{journal}{JHEP} \bibinfo{volume}{03} (\bibinfo{year}{2009})
  \bibinfo{pages}{054}. \DOIprefix\doi{10.1088/1126-6708/2009/03/054}.
  \href{http://arxiv.org/abs/0811.2067}{{\tt arXiv:0811.2067}}.
\bibitem[{Laenen et~al.(2011)Laenen, Magnea, Stavenga, and
  White}]{Laenen:2010uz}
\bibinfo{author}{E.~Laenen}, \bibinfo{author}{L.~Magnea},
  \bibinfo{author}{G.~Stavenga}, \bibinfo{author}{C.~D. White},
\newblock \bibinfo{title}{{Next-to-eikonal corrections to soft gluon radiation:
  a diagrammatic approach}},
\newblock \bibinfo{journal}{JHEP} \bibinfo{volume}{01} (\bibinfo{year}{2011})
  \bibinfo{pages}{141}. \DOIprefix\doi{10.1007/JHEP01(2011)141}.
  \href{http://arxiv.org/abs/1010.1860}{{\tt arXiv:1010.1860}}.
\bibitem[{Altinoluk et~al.(2014)Altinoluk, Armesto, Beuf, Martinez, and
  Salgado}]{Altinoluk:2014oxa}
\bibinfo{author}{T.~Altinoluk}, \bibinfo{author}{N.~Armesto},
  \bibinfo{author}{G.~Beuf}, \bibinfo{author}{M.~Martinez},
  \bibinfo{author}{C.~A. Salgado},
\newblock \bibinfo{title}{{Next-to-eikonal corrections in the CGC: gluon
  production and spin asymmetries in pA collisions}},
\newblock \bibinfo{journal}{JHEP} \bibinfo{volume}{07} (\bibinfo{year}{2014})
  \bibinfo{pages}{068}. \DOIprefix\doi{10.1007/JHEP07(2014)068}.
  \href{http://arxiv.org/abs/1404.2219}{{\tt arXiv:1404.2219}}.
\bibitem[{Altinoluk et~al.(2016)Altinoluk, Armesto, Beuf, and
  Moscoso}]{Altinoluk:2015gia}
\bibinfo{author}{T.~Altinoluk}, \bibinfo{author}{N.~Armesto},
  \bibinfo{author}{G.~Beuf}, \bibinfo{author}{A.~Moscoso},
\newblock \bibinfo{title}{{Next-to-next-to-eikonal corrections in the CGC}},
\newblock \bibinfo{journal}{JHEP} \bibinfo{volume}{01} (\bibinfo{year}{2016})
  \bibinfo{pages}{114}. \DOIprefix\doi{10.1007/JHEP01(2016)114}.
  \href{http://arxiv.org/abs/1505.01400}{{\tt arXiv:1505.01400}}.
\bibitem[{Altinoluk and Dumitru(2016)}]{Altinoluk:2015xuy}
\bibinfo{author}{T.~Altinoluk}, \bibinfo{author}{A.~Dumitru},
\newblock \bibinfo{title}{{Particle production in high-energy collisions beyond
  the shockwave limit}},
\newblock \bibinfo{journal}{Phys. Rev.} \bibinfo{volume}{D94}
  (\bibinfo{year}{2016}) \bibinfo{pages}{074032}.
  \DOIprefix\doi{10.1103/PhysRevD.94.074032}.
  \href{http://arxiv.org/abs/1512.00279}{{\tt arXiv:1512.00279}}.
\bibitem[{Agostini et~al.(2019)Agostini, Altinoluk, and
  Armesto}]{Agostini:2019avp}
\bibinfo{author}{P.~Agostini}, \bibinfo{author}{T.~Altinoluk},
  \bibinfo{author}{N.~Armesto},
\newblock \bibinfo{title}{{Non-eikonal corrections to multi-particle production
  in the Color Glass Condensate}},
\newblock \bibinfo{journal}{Eur. Phys. J.} \bibinfo{volume}{C79}
  (\bibinfo{year}{2019}) \bibinfo{pages}{600}.
  \DOIprefix\doi{10.1140/epjc/s10052-019-7097-5}.
  \href{http://arxiv.org/abs/1902.04483}{{\tt arXiv:1902.04483}}.
\bibitem[{Kowalski et~al.(2006)Kowalski, Motyka, and Watt}]{Kowalski:2006hc}
\bibinfo{author}{H.~Kowalski}, \bibinfo{author}{L.~Motyka},
  \bibinfo{author}{G.~Watt},
\newblock \bibinfo{title}{{Exclusive diffractive processes at HERA within the
  dipole picture}},
\newblock \bibinfo{journal}{Phys. Rev.} \bibinfo{volume}{D74}
  (\bibinfo{year}{2006}) \bibinfo{pages}{074016}.
  \DOIprefix\doi{10.1103/PhysRevD.74.074016}.
  \href{http://arxiv.org/abs/hep-ph/0606272}{{\tt arXiv:hep-ph/0606272}}.
\bibitem[{Kovner and Skokov(2018)}]{Kovner:2018azs}
\bibinfo{author}{A.~Kovner}, \bibinfo{author}{V.~V. Skokov},
\newblock \bibinfo{title}{{Bose enhancement, the Liouville effective action and
  the high multiplicity tail in p-A collisions}},
\newblock \bibinfo{journal}{Phys. Rev.} \bibinfo{volume}{D98}
  (\bibinfo{year}{2018}) \bibinfo{pages}{014004}.
  \DOIprefix\doi{10.1103/PhysRevD.98.014004}.
  \href{http://arxiv.org/abs/1805.09296}{{\tt arXiv:1805.09296}}.

\end{thebibliography}

\end{document}